\documentclass{gGAF2e}
\usepackage{graphicx,natbib,bm,url,color}
\graphicspath{{./fig/}{./png/}}

\newcommand{\Lu}{L_{\rm u}}
\newcommand{\Ld}{L_{\rm d}}
\newcommand{\Hd}{H_{\rm d}}

\newcommand{\rd}{{\rho_{\rm d}}}

\newcommand{\csu}{c_{\rm su}}
\newcommand{\csd}{c_{\rm sd}}

\newcommand{\emfA}{{\cal E}_{0y}^{\rm A}}
\newcommand{\emfB}{{\cal E}_{0y}^{\rm B}}
\newcommand{\FTH}{{\cal F}^{\rm TH}(x;x_1,x_2)}

\begin{document}
\jvol{00} \jnum{00} \jyear{2018} 

\markboth{Singh et al.}{Magnetically induced perturbations on the $f$-mode}

\articletype{GAFD Special issue on ``Physics and Algorithms of the Pencil Code''}


\newcommand{\EQ}{\begin{equation}}
\newcommand{\EN}{\end{equation}}
\newcommand{\EQA}{\begin{eqnarray}}
\newcommand{\ENA}{\end{eqnarray}}
\newcommand{\eq}[1]{(\ref{#1})}
\newcommand{\EEq}[1]{Equation~(\ref{#1})}
\newcommand{\Eq}[1]{equation~(\ref{#1})}
\newcommand{\Eqs}[2]{equations~(\ref{#1}) and~(\ref{#2})}
\newcommand{\EEqs}[2]{equations~(\ref{#1}) and~(\ref{#2})}
\newcommand{\Eqss}[2]{equations~(\ref{#1})--(\ref{#2})}
\newcommand{\eqs}[2]{(\ref{#1}) and~(\ref{#2})}
\newcommand{\App}[1]{appendix~\ref{#1}}

\newcommand{\Sec}[1]{section~\ref{#1}}
\newcommand{\Secs}[2]{sections~\ref{#1} and \ref{#2}}

\newcommand{\Fig}[1]{figure~\ref{#1}}
\newcommand{\FFig}[1]{Figure~\ref{#1}}
\newcommand{\Figp}[2]{figure~\ref{#1}({#2})}
\newcommand{\Figsp}[3]{figures~\ref{#1}({#2}) and ({#3})}
\newcommand{\Figssp}[3]{figures~\ref{#1}({#2})--({#3})}
\newcommand{\Figs}[2]{figures~\ref{#1} and \ref{#2}}
\newcommand{\Figss}[2]{figures~\ref{#1}--\ref{#2}}
\newcommand{\Tab}[1]{table~\ref{#1}}
\newcommand{\Tabs}[2]{tables~\ref{#1} and \ref{#2}}


\newcommand{\bra}[1]{\langle #1\rangle}
\newcommand{\bbra}[1]{\left\langle #1\right\rangle}
\newcommand{\mean}[1]{\overline #1}
\newcommand{\meanrho}{\overline{\rho}}
\newcommand{\meanPhi}{\overline{\Phi}}
\newcommand{\tildeFFFF}{\tilde{\mbox{\boldmath ${\cal F}$}}{}}{}
\newcommand{\hatFFFF}{\hat{\mbox{\boldmath ${\cal F}$}}{}}{}
\newcommand{\meanFFFF}{\overline{\mbox{\boldmath ${\cal F}$}}{}}{}
\newcommand{\meanemf}{\overline{\cal E} {}}
\newcommand{\meanAAAA}{\overline{\mbox{\boldmath ${\mathsf A}$}} {}}
\newcommand{\meanSSSS}{\overline{\mbox{\boldmath ${\mathsf S}$}} {}}
\newcommand{\meanAAA}{\overline{\mathsf{A}}}
\newcommand{\meanSSS}{\overline{\mathsf{S}}}
\newcommand{\meanuu}{\overline{\mbox{\boldmath $u$}}{}}{}
\newcommand{\meanoo}{\overline{\mbox{\boldmath $\omega$}}{}}{}
\newcommand{\meanEMF}{\overline{\mbox{\boldmath ${\cal E}$}}{}}{}
\newcommand{\meanEEEE}{\overline{\mbox{\boldmath ${\cal E}$}}{}}{}
\newcommand{\meanuxB}{\overline{\mbox{\boldmath $\delta u\times \delta B$}}{}}{}
\newcommand{\meanJB}{\overline{\mbox{\boldmath $J\cdot B$}}{}}{}
\newcommand{\meanAB}{\overline{\mbox{\boldmath $A\cdot B$}}{}}{}
\newcommand{\meanjb}{\overline{\mbox{\boldmath $j\cdot b$}}{}}{}
\newcommand{\meanEE}{\overline{\mbox{\boldmath $E$}}{}}{}
\newcommand{\meanFFf}{\overline{\mbox{\boldmath $F$}}_{\rm f}{}}{}
\newcommand{\meanFFm}{\overline{\mbox{\boldmath $F$}}_{\rm m}{}}{}
\newcommand{\hatFF}{\hat{\mbox{\boldmath $F$}}{}}{}
\newcommand{\meanGG}{\overline{\mbox{\boldmath $G$}}{}}{}
\newcommand{\meanHH}{\overline{\mbox{\boldmath $H$}}{}}{}
\newcommand{\tildeGG}{\tilde{\mbox{\boldmath $G$}}{}}{}
\newcommand{\meanCC}{\overline{\mbox{\boldmath $C$}}{}}{}
\newcommand{\meanKK}{\overline{\mbox{\boldmath $K$}}{}}{}
\newcommand{\meanUU}{\overline{\bm{U}}}
\newcommand{\meanVV}{\overline{\bm{V}}}
\newcommand{\meanWW}{\overline{\bm{W}}}
\newcommand{\meanQQ}{\overline{\mbox{\boldmath $Q$}}{}}{}
\newcommand{\meana}{\overline{a}}
\newcommand{\meanA}{\overline{A}}
\newcommand{\meanB}{\overline{B}}
\newcommand{\meanE}{\overline{E}}
\newcommand{\meanK}{\overline{K}}
\newcommand{\meanb}{\tilde{b}}
\newcommand{\meanj}{\tilde{j}}
\newcommand{\meanC}{\overline{C}}
\newcommand{\meanD}{\overline{D}}
\newcommand{\meanF}{\overline{F}}
\newcommand{\meanFm}{\overline{F}_{\rm m}}
\newcommand{\meanFf}{\overline{F}_{\rm f}}
\newcommand{\meanh}{\overline{h}}
\newcommand{\meanhm}{\overline{h}_{\rm m}}
\newcommand{\meanhf}{\overline{h}_{\rm f}}
\newcommand{\meanG}{\overline{G}}
\newcommand{\meanGGG}{\overline{\cal G}}
\newcommand{\meanH}{\overline{H}}
\newcommand{\meanU}{\overline{U}}
\newcommand{\meanR}{\overline{R}}
\newcommand{\meanJ}{\overline{J}}
\newcommand{\means}{\overline{s}}
\newcommand{\meanp}{\overline{p}}
\newcommand{\meanP}{\overline{P}}
\newcommand{\meanS}{\overline{S}}
\newcommand{\meanT}{\overline{T}}
\newcommand{\meanW}{\overline{W}}
\newcommand{\meanQ}{\overline{Q}}
\newcommand{\meanEEE}{\overline{\cal E}}
\newcommand{\meanFFF}{\overline{\cal F}}
\newcommand{\meanHHH}{\overline{\cal H}}
\newcommand{\meanRRR}{\overline{\cal R}}
\newcommand{\hatg}{\hat{g}}
\newcommand{\hatk}{\hat{k}}
\newcommand{\hatkk}{\hat{\bm{k}}}
\newcommand{\hatAA}{\hat{\bm{A}}}
\newcommand{\hatBB}{\hat{\bm{B}}}
\newcommand{\hatJJ}{\hat{\mbox{\boldmath $J$}}{}}{}
\newcommand{\hatOO}{\hat{\bm{\Omega}}}
\newcommand{\hatEMF}{\hat{\mbox{\boldmath ${\cal E}$}}{}}{}
\newcommand{\Rgas}{{\cal R}}{}
\newcommand{\emf}{{\cal E}}{}
\newcommand{\hatA}{\hat{A}}
\newcommand{\hatB}{\hat{B}}
\newcommand{\hatJ}{\hat{J}}
\newcommand{\hatU}{\hat{U}}
\newcommand{\alphaK}{\alpha_{\rm K}}
\newcommand{\alphaM}{\alpha_{\rm M}}
\newcommand{\alphaSS}{\alpha_{\rm SS}}
%
%
\newcommand{\tildeB}{\tilde{B}}
\newcommand{\tildeJ}{\tilde{J}}
\newcommand{\tildeemf}{\tilde{\cal E}}
\newcommand{\teps}{\tilde{\epsilon} {}}
\newcommand{\tkapz}{\tilde{\kappa_0}}
\newcommand{\Oh}{\hat{\Omega}}
\newcommand{\zh}{\hat{z}}
\newcommand{\PC}{{\sc Pencil Code}~}
%
%
\newcommand{\pphi}{\hat{\bm{\phi}}}
\newcommand{\ppom}{\bm{\hat{\varpi}}}
\newcommand{\eee}{\hat{\mbox{\boldmath $e$}} {}}
\newcommand{\nnn}{\hat{\bm{n}}}
\newcommand{\rrr}{\hat{\mbox{\boldmath $r$}} {}}
\newcommand{\vvv}{\hat{\mbox{\boldmath $v$}} {}}
\newcommand{\xxx}{\hat{\mbox{\boldmath $x$}} {}}
\newcommand{\yyy}{\hat{\mbox{\boldmath $y$}} {}}
\newcommand{\zzz}{\hat{\mbox{\boldmath $z$}} {}}
\newcommand{\ttt}{\hat{\mbox{\boldmath $\theta$}} {}}
\newcommand{\OOO}{\hat{\mbox{\boldmath $\Omega$}} {}}
\newcommand{\ooo}{\hat{\mbox{\boldmath $\omega$}} {}}
\newcommand{\BBBB}{\hat{\mbox{\boldmath $B$}} {}}
\newcommand{\BBhat}{\hat{\bm{B}}}
\newcommand{\meanAA}{{\overline{\bm{A}}}}
\newcommand{\meanBB}{{\overline{\bm{B}}}}
\newcommand{\meanJJ}{{\overline{\bm{J}}}}
\newcommand{\meanFF}{{\overline{\bm{F}}}}
\newcommand{\meanBBhat}{\hat{\overline{\bm{B}}}}
\newcommand{\meanJJhat}{\hat{\overline{\bm{J}}}}
\newcommand{\meanBhat}{\hat{\overline{B}}}
\newcommand{\meanJhat}{\hat{\overline{J}}}
\newcommand{\Bhat}{\hat{B}}
\newcommand{\Bh}{\hat{B}}
%
%
\newcommand{\nullvector}{{\bf0}}
\newcommand{\gggg}{\mbox{\boldmath $g$} {}}
\newcommand{\ddd}{\mbox{\boldmath $d$} {}}
\newcommand{\yy}{\mbox{\boldmath $y$} {}}
\newcommand{\zz}{\mbox{\boldmath $z$} {}}
\newcommand{\vv}{\mbox{\boldmath $v$} {}}
\newcommand{\ww}{\mbox{\boldmath $w$} {}}
\newcommand{\mm}{\mbox{\boldmath $m$} {}}
\newcommand{\PP}{\mbox{\boldmath $P$} {}}
\newcommand{\bp}{\mbox{\boldmath $p$} {}}
\newcommand{\II}{\mbox{\boldmath $I$} {}}
\newcommand{\RR}{\mbox{\boldmath $R$} {}}
\newcommand{\kk}{\bm{k}}
\newcommand{\KK}{\bm{K}}
\newcommand{\pp}{\bm{p}}
\newcommand{\qq}{\bm{q}}
\newcommand{\xx}{\bm{x}}
\newcommand{\XX}{\bm{X}}
\newcommand{\aaaa}{\bm{a}}
\newcommand{\jj}{\bm{j}}
\newcommand{\rr}{\bm{r}}
\newcommand{\grav}{\bm{g}}
\newcommand{\BB}{\bm{B}}
\newcommand{\CC}{\bm{C}}
\newcommand{\EE}{\bm{E}}
\newcommand{\JJ}{\bm{J}}
\newcommand{\oo}{\bm{\omega}}
\newcommand{\AAA}{\bm{A}}
\newcommand{\HH}{\bm{H}}
\newcommand{\MM}{\bm{M}}
\newcommand{\UU}{\bm{U}}
\newcommand{\FF}{\bm{F}}
\def\os{\omega_{\rm c}}
\def\oA{\omega_{\rm A}}
\def\oAy{\omega_{{\rm A}y}}
\def\oAz{\omega_{{\rm A}z}}
\newcommand{\vvA}{\bm{v}_{\rm A}}
\newcommand{\uu}{\bm{u}}
\newcommand{\uud}{\bm{u}_{\rm d}}
\newcommand{\rhod}{\rho_{\rm d}}
\newcommand{\sss}{\mbox{\boldmath $s$} {}}
\newcommand{\SSS}{\mbox{\boldmath $S$} {}}
\newcommand{\ee}{\mbox{\boldmath $e$} {}}
\newcommand{\nn}{\mbox{\boldmath $n$} {}}
\newcommand{\ff}{\mbox{\boldmath $f$} {}}
\newcommand{\hh}{\mbox{\boldmath $h$} {}}
\newcommand{\EEE}{\mbox{\boldmath ${\cal E}$} {}}
\newcommand{\FFF}{\mbox{\boldmath ${\cal F}$} {}}
\newcommand{\TT}{{\bm{T}}}
\newcommand{\nab}{{\bm{\nabla}}}
\newcommand{\GG}{\mbox{\boldmath $G$} {}}
\newcommand{\WW}{\mbox{\boldmath $W$} {}}
\newcommand{\QQ}{\mbox{\boldmath $Q$} {}}
\newcommand{\nabD}{\nabla_{\rm D}}
\newcommand{\nabel}{\nabla_{\rm el}}
\newcommand{\nabad}{\nabla_{\rm ad}}
\newcommand{\nabrad}{\nabla_{\rm rad}}
\newcommand{\nabk}{\mbox{\boldmath $\nabla$}_{\!\kk\,}}
\newcommand{\OO}{\bm{\Omega}}
\newcommand{\ppsi}{\mbox{\boldmath $\psi$} {}}
\newcommand{\pom}{\mbox{\boldmath $\varpi$} {}}
\newcommand{\ttau}{\mbox{\boldmath $\tau$} {}}
\newcommand{\LL}{\mbox{\boldmath $\Lambda$} {}}
\newcommand{\mmu}{\mbox{\boldmath $\mu$} {}}
\newcommand{\xxi}{\mbox{\boldmath $\xi$} {}}
\newcommand{\ddelta}{\mbox{\boldmath $\delta$} {}}
\newcommand{\ggamma}{\mbox{\boldmath $\gamma$} {}}
\newcommand{\kkappa}{\mbox{\boldmath $\kappa$} {}}
\newcommand{\llambda}{\mbox{\boldmath $\lambda$} {}}
\newcommand{\pomega}{\mbox{\boldmath $\varpi$} {}}
\newcommand{\PPsi}{\mbox{\boldmath $\Psi$} {}}
%
%
\newcommand{\DDDD}{\mbox{\boldmath ${\sf D}$} {}}
\newcommand{\IIII}{\mbox{\boldmath ${\sf I}$} {}}
\newcommand{\LLLL}{\mbox{\boldmath ${\sf L}$} {}}
\newcommand{\MMMM}{\mbox{\boldmath ${\sf M}$} {}}
\newcommand{\NNNN}{\mbox{\boldmath ${\sf N}$} {}}
\newcommand{\PPPP}{\mbox{\boldmath ${\sf P}$} {}}
\newcommand{\QQQQ}{\mbox{\boldmath ${\sf Q}$} {}}
\newcommand{\RRRR}{\mbox{\boldmath ${\sf R}$} {}}
\newcommand{\SSSS}{\mbox{\boldmath ${\sf S}$} {}}
\newcommand{\BBBBB}{\mbox{\boldmath ${\sf B}$} {}}
\newcommand{\tAAAA}{\tilde{\mbox{\boldmath ${\sf A}$}} {}}
\newcommand{\tDDDD}{\tilde{\mbox{\boldmath ${\sf D}$}} {}}
\newcommand{\tRRRR}{\tilde{\mbox{\boldmath ${\sf R}$}} {}}
\newcommand{\tQQQQ}{\tilde{\mbox{\boldmath ${\sf Q}$}} {}}
\newcommand{\AAAA}{\mbox{\boldmath ${\cal A}$} {}}
\newcommand{\BBB}{\mbox{\boldmath ${\cal B}$} {}}
\newcommand{\EMF}{\mbox{\boldmath ${\cal E}$} {}}
\newcommand{\GGG}{\mbox{\boldmath ${\cal G}$} {}}
\newcommand{\HHH}{\mbox{\boldmath ${\cal H}$} {}}
\newcommand{\meanQQQ}{\overline{\mbox{\boldmath ${\cal Q}$}} {}}
\newcommand{\GGGG}{{\bf G} {}}
%
%
\newcommand{\ii}{{\rm i}}
\newcommand{\erf}{{\rm erf}}
\newcommand{\grad}{{\rm grad} \, {}}
\newcommand{\curl}{{\rm curl} \, {}}
\newcommand{\dive}{{\rm div}  \, {}}
\newcommand{\Dive}{{\rm Div}  \, {}}
\newcommand{\diag}{{\rm diag}  \, {}}
\newcommand{\asin}{{\rm asin}  \, {}}
\newcommand{\sgn}{{\rm sgn}  \, {}}
\newcommand{\meanDD}{{\overline{\rm D}} {}}
\newcommand{\DD}{{\rm D} {}}
\newcommand{\DDD}{{\cal D} {}}
\newcommand{\QQQ}{{\cal Q}}
\newcommand{\dd}{{\rm d} {}}
\newcommand{\dV}{\,{\rm d}V {}}
\newcommand{\dS}{\,{\rm d}{{\bm{S}}} {}}
\newcommand{\const}{{\rm const}  {}}
\newcommand{\CR}{{\rm CR}}
\def\degr{\hbox{$^\circ$}}
\def\la{\mathrel{\mathchoice {\vcenter{\offinterlineskip\halign{\hfil
$\displaystyle##$\hfil\cr<\cr\sim\cr}}}
{\vcenter{\offinterlineskip\halign{\hfil$\textstyle##$\hfil\cr<\cr\sim\cr}}}
{\vcenter{\offinterlineskip\halign{\hfil$\scriptstyle##$\hfil\cr<\cr\sim\cr}}}
{\vcenter{\offinterlineskip\halign{\hfil$\scriptscriptstyle##$\hfil\cr<\cr\sim\cr}}}}}
\def\ga{\mathrel{\mathchoice {\vcenter{\offinterlineskip\halign{\hfil
$\displaystyle##$\hfil\cr>\cr\sim\cr}}}
{\vcenter{\offinterlineskip\halign{\hfil$\textstyle##$\hfil\cr>\cr\sim\cr}}}
{\vcenter{\offinterlineskip\halign{\hfil$\scriptstyle##$\hfil\cr>\cr\sim\cr}}}
{\vcenter{\offinterlineskip\halign{\hfil$\scriptscriptstyle##$\hfil\cr>\cr\sim\cr}}}}}
%
%
\def\H{\mbox{\rm H}}
\def\pH{\mbox{\rm pH}}
\def\Ta{\mbox{\rm Ta}}
\def\Ra{\mbox{\rm Ra}}
\def\Rat{\mbox{\rm Ra}_{\rm turb}}
\def\Prt{\mbox{\rm Pr}_{\rm turb}}
\def\Ma{\mbox{\rm Ma}}
\def\Co{\mbox{\rm Co}}
\def\Ro{\mbox{\rm Ro}}
\def\Ri{\mbox{\rm Ri}}
\def\CoM{\mbox{\rm Co}_{\rm M}}
\def\Sh{\mbox{\rm Sh}}
\def\St{\mbox{\rm St}}
\def\Roo{\mbox{\rm Ro}^{-1}}
\def\Rooo{\mbox{\rm Ro}^{-2}}
\def\Pra{\mbox{\rm Pr}}
\def\Prm{\mbox{\rm Pr}_{\rm m}}
\def\Pran{\mbox{\rm Pr}}
\def\Sc{\mbox{\rm Sc}}
\def\Tr{\mbox{\rm Tr}}
\def\Da{\mbox{\rm Da}}
\def\Ka{\mbox{\rm Ka}}
\def\RM{\mbox{\rm RM}}
\def\Pm{\mbox{\rm Pr}_M}
\def\Rm{R_{\rm m}}
\def\Dt{D_{\rm t}}
\def\DT{D_{\rm T}}
\def\Rmc{R_{\rm m,{\rm crit}}}
\def\Rey{\mbox{\rm Re}}
\def\Imag{\mbox{\rm Im}}
\def\Pe{\mbox{\rm Pe}}
\def\Co{\mbox{\rm Co}}
\def\Gr{\mbox{\rm Gr}}
\def\Lu{\mbox{\rm Lu}}
\def\tauc{\tau_{\rm c}}
\def\tauA{\tau_{\rm A}}
\def\taueff{\tau_{\rm eff}}
\def\HP{H_{\!P}}
\def\cP{c_{P}}
\def\cV{c_{V}}
\def\alpK{\alpha_{\rm K}}
\def\alpM{\alpha_{\rm M}}
\def\EEK{{\cal E}_{\rm K}}
\def\EEM{{\cal E}_{\rm M}}
\def\HHK{{\cal H}_{\rm K}}
\def\HHM{{\cal H}_{\rm M}}
\def\EK{E_{\rm K}}
\def\EM{E_{\rm M}}
\def\EKM{E_{\rm K/M}}
\def\ET{E_{\rm T}}
\def\WL{W_{\rm L}}
\def\WC{W_{\rm C}}
\def\cmag{c_{\rm m}}
\def\cmso{c_{\rm ms}}
\def\csz{c_{\rm s0}}
\def\Teff{T_{\rm eff}}
\def\cp{c_{\rm p}}
\def\cv{c_{\rm v}}
\def\Rs{R_{\odot}}
\def\cs{c_{\rm s}}
\def\xiM{\xi_{\rm M}}
\def\xiK{\xi_{\rm K}}
\def\xiKM{\xi_{\rm K/M}}
\def\Hpz{H_{\rm p0}}
\def\Hrz{H_{\rho 0}}
\def\qpz{q_{\rm p0}}
\def\qp{q_{\rm p}}
\def\betaK{\beta_{\rm K}}
\def\betaM{\beta_{\rm M}}
\def\betaKM{\beta_{\rm K/M}}
\def\betap{\beta_{\rm p}}
\def\betamin{\beta_{\min}}
\def\betastar{\beta_{\star}}
\def\Peff{{\cal P}_{\rm eff}}
\def\Pmin{{\cal P}_{\rm min}}
\def\pturbz{p_{\rm turb}^{(0)}}
\def\pturb{p_{\rm turb}}
\def\peff{p_{\rm eff}}
\def\qs{q_{\rm s}}
\def\qe{q_{\rm e}}
\def\qg{q_{\rm g}}
\def\qa{q_{\rm a}}
\def\pt{p_{\rm t}}
\def\ptz{p_{\rm t0}}
\def\vA{v_{\rm A}}
\def\vAy{v_{{\rm A}y}}
\def\vAZ{v_{{\rm A}z}}
\def\vAz{v_{\rm A0}}
\def\hf{h_{\rm f}}
\def\hm{h_{\rm m}}
\def\hrms{h_{\rm rms}}
\def\kmean{k_{\rm m}}
\def\lf{l_{\rm f}}
\def\kone{k_1}
\def\kd{k_{\rm d}}
\def\kf{k_{\rm f}}
\def\kp{k_{\rm p}}
\def\Hp{H_{\rm p}}
\def\HM{H_{\rm M}}
\def\EM{E_{\rm M}}
\def\Ekin{E_{\rm kin}}
\def\kB{k_{\rm B}}
\def\sigmaSB{\sigma_{\rm SB}}
\def\kappaf{\kappa_{\rm f}}
\def\epsf{\epsilon_{\rm f}}
\def\epsfz{\epsilon_{\rm f0}}
\def\epsK{\epsilon_{\it K}}
\def\epsM{\epsilon_{\it M}}
\def\epsC{\epsilon_{\it C}}
\def\epsT{\epsilon_{\it T}}
\def\betK{\beta_{\it K}}
\def\betM{\beta_{\it M}}
\def\xid{\xi_{\rm D}}
\def\NA{N_{\rm A}}
\def\kB{k_{\rm B}}
\def\kK{k_{\it K}}
\def\kM{k_{\it M}}
\def\kmax{k_{\rm max}}
\def\kom{k_{\omega}}
\def\Ff{F_{\rm f}}
\def\Fm{F_{\rm m}}
\def\Hf{H_{\rm f}}
\def\Hm{H_{\rm m}}
\def\vArms{v_{\rm A,rms}}
\def\Brms{B_{\rm rms}}
\def\Jrms{J_{\rm rms}}
\def\Urms{U_{\rm rms}}
\def\urms{u_{\rm rms}}
\def\wrms{w_{\rm rms}}
\def\orms{\omega_{\rm rms}}
\def\omms{\omega_{\rm ms}}
\def\brms{b_{\rm rms}}
\def\uref{u_{\rm ref}}
\def\yN{y_{\rm N}}
\def\yL{y_{\rm L}}
\def\kappaBB{\kappa_{\rm BB}}
\def\kappaB{\kappa_{\rm B}}
\def\kappaOO{\kappa_{\Omega\Omega}}
\def\kappaO{\kappa_{\Omega}}
\def\kappah{\kappa_{\rm h}}
\def\kappat{\kappa_{\rm t}}
\def\kappaT{\kappa_{\rm T}}
\def\kappatz{\kappa_{\rm t0}}
\def\mut{\mu_{\rm t}}
\def\nut{\nu_{\rm t}}
\def\nuM{\nu_{\rm M}}
\def\nuT{\nu_{\rm T}}
\def\calKt{{\cal K}_{\rm t}}
\def\calKtz{{\cal K}_{\rm t0}}
\def\chit{\chi_{\rm t}}
\def\chitz{\chi_{\rm t0}}
\def\etat{\eta_{\rm t}}
\def\etatz{\eta_{\rm t0}}
\def\etatz{\eta_{\rm t0}}
\def\etaTz{\eta_{\rm T0}}
\def\nutz{\nu_{\rm t0}}
\def\etaT{\eta_{\rm T}}
\def\mut{\mu_{\rm t}}
\def\muT{\mu_{\rm T}}
\def\uT{\mu{\rm T}}
\def\BBeq{|\BB|/B_{\rm eq}}
\def\Beq{B_{\rm eq}}
\def\Beqz{B_{\rm eq0}}
\def\tautd{\tau_{\rm td}}
\def\tauto{\tau_{\rm to}}
\newcommand{\ea}{{\rm et al.\ }}
\newcommand{\eaa}{{\rm et al.\ }}
\def\half{{\textstyle{1\over2}}}
\def\hhalf{{\scriptstyle{1\over2}}}
\def\threehalf{{\textstyle{3\over2}}}
\def\threequarter{{\textstyle{3\over4}}}
\def\sevenhalf{{\textstyle{7\over2}}}
\def\onethird{{\textstyle{1\over3}}}
\def\onesixth{{\textstyle{1\over6}}}
\def\twothird{{\textstyle{2\over3}}}
\def\fivethird{{\textstyle{5\over3}}}
\def\onenineth{{\textstyle{1\over9}}}
\def\fourthird{{\textstyle{4\over3}}}
\def\quarter{{\textstyle{1\over4}}}
\newcommand{\W}{\,{\rm W}}
\newcommand{\V}{\,{\rm V}}
\newcommand{\kV}{\,{\rm kV}}
\newcommand{\T}{\,{\rm T}}
\newcommand{\G}{\,{\rm G}}
\newcommand{\sfu}{\,{\rm sfu}}
\newcommand{\Jy}{\,{\rm Jy}}
\newcommand{\Hz}{\,{\rm Hz}}
\newcommand{\nHz}{\,{\rm nHz}}
\newcommand{\GHz}{\,{\rm GHz}}
\newcommand{\kHz}{\,{\rm kHz}}
\newcommand{\kG}{\,{\rm kG}}
\newcommand{\K}{\,{\rm K}}
\newcommand{\g}{\,{\rm g}}
\newcommand{\s}{\,{\rm s}}
\newcommand{\um}{\,\mu{\rm m}}
\newcommand{\ms}{\,{\rm ms}}
\newcommand{\mpers}{\,{\rm m/s}}
\newcommand{\ks}{\,{\rm ks}}
\newcommand{\cm}{\,{\rm cm}}
\newcommand{\mV}{\,{\rm mV}}
\newcommand{\cmcube}{\,{\rm cm^{-3}}}
\newcommand{\nm}{\,{\rm nm}}
\newcommand{\m}{\,{\rm m}}
\newcommand{\km}{\,{\rm km}}
\newcommand{\msec}{\,{\rm ms}}
\newcommand{\cms}{\,{\rm cm/s}}
\newcommand{\kilom}{\,{\rm km}}
\newcommand{\kms}{\,{\rm km/s}}
\newcommand{\gpercc}{\,{\rm g/cm}^3}
\newcommand{\kg}{\,{\rm kg}}
\newcommand{\ug}{\,\mu{\rm g}}
\newcommand{\kW}{\,{\rm kW}}
\newcommand{\MW}{\,{\rm MW}}
\newcommand{\Mm}{\,{\rm Mm}}
\newcommand{\Mx}{\,{\rm Mx}}
\newcommand{\pc}{\,{\rm pc}}
\newcommand{\kpc}{\,{\rm kpc}}
\newcommand{\yr}{\,{\rm yr}}
\newcommand{\Myr}{\,{\rm Myr}}
\newcommand{\Gyr}{\,{\rm Gyr}}
\newcommand{\erg}{\,{\rm erg}}
\newcommand{\mol}{\,{\rm mol}}
\newcommand{\dyn}{\,{\rm dyn}}
\newcommand{\J}{\,{\rm J}}
\newcommand{\AU}{\,{\rm AU}}
\newcommand{\A}{\,{\rm A}}
\newcommand{\etal}{et al.}
\newcommand{\Hminus}{{\rm H}^{-}}
\newcommand{\Mdot}{\dot{M}}
\newcommand{\MSun}{M_\odot}
%
%
%
\newcommand{\arXiv}[3]{, ``#2,'' arXiv:#3 (#1).}
\newcommand{\ypasj}[5]{, ``#5,'' {\em Publ. Astron. Soc. Jap.\ }{\bf #2}, #3-#4 (#1).}
\newcommand{\ypasjN}[4]{, ``#4,'' {\em Publ. Astron. Soc. Jap.\ }{\bf #2}, #3 (#1).}
\newcommand{\ynjpN}[4]{, ``#4,'' {\em New J. Phys.\ }{\bf #2}, #3 (#1).}
\newcommand{\yphl}[5]{, ``#5,'' {\em Phys.\ Lett.\ }{\bf #2}, #3--#4 (#1).}
\newcommand{\yan}[5]{, ``#5,'' {\em Astron.\ Nachr.\ }{\bf #2}, #3--#4 (#1).}
\newcommand{\yanS}[5]{, ``#5'' {\em Astron.\ Nachr.\ }{\bf #2}, #3--#4 (#1).}
\newcommand{\yact}[5]{, ``#5,'' {\em Acta Astron.\ }{\bf #2}, #3--#4 (#1).}
\newcommand{\ypnas}[5]{, ``#5,'' {\em Proc.\ Nat.\ Acad.\ Sci.\ }{\bf #2}, #3--#4 (#1).}
\newcommand{\ylrsp}[4]{, ``#4,'' {\em Liv.\ Rev.\ Solar Phys.\ }{\bf #2}, #3 (#1).}
\newcommand{\yana}[5]{, ``#5,'' {\em Astron.\ Astrophys.\ }{\bf #2}, #3--#4 (#1).}
\newcommand{\dana}[3]{, ``#2,'' {\em Astron.\ Astrophys.}, in press, DOI:#3 (#1).}
\newcommand{\pana}[3]{, ``#2,'' {\em Astron.\ Astrophys.}, in press, arXiv:#3 (#1).}
\newcommand{\sana}[2]{, ``#2,'' {\em Astron.\ Astrophys.}, submitted (#1).}
\newcommand{\tana}[3]{, ``#2,'' {\em Astron.\ Astrophys.}, to be submitted, \url{#3} (#1).}
\newcommand{\yanaN}[4]{, ``#4,'' {\em Astron.\ Astrophys.\ }{\bf #2}, #3 (#1).}
\newcommand{\yanaS}[5]{, ``#5'' {\em Astron.\ Astrophys.\ }{\bf #2}, #3--#4 (#1).}
\newcommand{\yanas}[5]{, ``#5,'' {\em Astron.\ Astrophys.\ Suppl.\ }{\bf #2}, #3--#4 (#1).}
\newcommand{\yanal}[5]{, ``#5,'' {\em Astron.\ Astrophys.\ Lett.\ }{\bf #2}, #3--#4 (#1).}
\newcommand{\yass}[5]{, ``#5,'' {\em Astrophys.\ Spa.\ Sci.\ }{\bf #2}, #3--#4 (#1).}
\newcommand{\yssr}[5]{, ``#5,'' {\em Spa.\ Sci.\ Rev.\ }{\bf #2}, #3--#4 (#1).}
\newcommand{\ysci}[5]{, ``#5,'' {\em Science }{\bf #2}, #3--#4 (#1).}
\newcommand{\ysciS}[4]{, ``#4,'' {\em Science }{\bf #2}, #3 (#1).}
\newcommand{\ysph}[5]{, ``#5,'' {\em Solar Phys.\ }{\bf #2}, #3--#4 (#1).}
\newcommand{\yjetp}[5]{, ``#5,'' {\em Sov.\ Phys.\ JETP }{\bf #2}, #3--#4 (#1).}
\newcommand{\yspd}[5]{, ``#5,'' {\em Sov.\ Phys.\ Dokl.\ }{\bf #2}, #3--#4 (#1).}
\newcommand{\yspdN}[4]{, ``#4,'' {\em Sov.\ Phys.\ Dokl.\ }{\bf #2}, #3 (#1).}
\newcommand{\ysov}[5]{, ``#5,'' {\em Sov.\ Astron.\ }{\bf #2}, #3--#4 (#1).}
\newcommand{\ysovl}[5]{, ``#5,'' {\em Sov.\ Astron.\ Lett.\ }{\bf #2}, #3--#4 (#1).}
\newcommand{\ymn}[5]{, ``#5,'' {\em Month. Not. Roy.\ Astron.\ Soc.\ }{\bf #2}, #3--#4 (#1).}
\newcommand{\dmn}[3]{, ``#3,'' {\em Month. Not. Roy.\ Astron.\ Soc.\ } DOI:#2 (#1).}
\newcommand{\yqjras}[5]{, ``#5,'' {\em Quart.\ J.\ Roy.\ Astron.\ Soc.\ }{\bf #2}, #3--#4 (#1).}
\newcommand{\ynat}[5]{, ``#5,'' {\em Nature }{\bf #2}, #3--#4 (#1).}
\newcommand{\san}[3]{, ``#2,'' {\em Astron.\ Nachr.\ } submitted, arXiv:#3 (#1).}
\newcommand{\sjfm}[3]{, ``#2,'' {\em J.\ Fluid Mech.\ } submitted, arXiv:#3 (#1).}
\newcommand{\pjfm}[2]{, ``#2,'' {\em J.\ Fluid Mech.\ } in press (#1).}
\newcommand{\yjfm}[5]{, ``#5,'' {\em J.\ Fluid Mech.\ }{\bf #2}, #3--#4 (#1).}
\newcommand{\ypr}[5]{, ``#5,'' {\em Phys.\ Rev.\ }{\bf #2}, #3--#4 (#1).}
\newcommand{\yprd}[5]{, ``#5,'' {\em Phys.\ Rev.\ D }{\bf #2}, #3--#4 (#1).}
\newcommand{\sprd}[3]{, ``#3,'' {\em Phys.\ Rev.\ D}, submitted, arXiv:#2 (#1).}
\newcommand{\epre}[5]{, ``#5,'' {\em Phys.\ Rev.\ E }{\bf #2}, #3--#4 (#1).}
\newcommand{\yprdN}[4]{, ``#4,'' {\em Phys.\ Rev.\ D }{\bf #2}, #3 (#1).}
\newcommand{\ypre}[5]{, ``#5,'' {\em Phys.\ Rev.\ E }{\bf #2}, #3--#4 (#1).}
\newcommand{\ypreN}[4]{, ``#4,'' {\em Phys.\ Rev.\ E }{\bf #2}, #3 (#1).}
\newcommand{\ypreNS}[4]{, ``#4'' {\em Phys.\ Rev.\ E }{\bf #2}, #3 (#1).}
\newcommand{\yprl}[5]{, ``#5,'' {\em Phys.\ Rev.\ Lett.\ }{\bf #2}, #3--#4 (#1).}
\newcommand{\yprlN}[4]{, ``#4,'' {\em Phys.\ Rev.\ Lett.\ }{\bf #2}, #3 (#1).}
\newcommand{\yepl}[5]{, ``#5,'' {\em Europhys.\ Lett.\ }{\bf #2}, #3--#4 (#1).}
\newcommand{\pcsf}[2]{, ``#2,'' {\em Chaos, Solitons \& Fractals} in press (#1).}
\newcommand{\ycsf}[5]{, ``#5,'' {\em Chaos, Solitons \& Fractals}{\bf #2}, #3--#4 (#1).}
\newcommand{\yprs}[5]{, ``#5,'' {\em Proc.\ Roy.\ Soc.\ Lond.\ }{\bf #2}, #3--#4 (#1).}
\newcommand{\yptrs}[5]{, ``#5,'' {\em Phil.\ Trans.\ Roy.\ Soc.\ }{\bf #2}, #3--#4 (#1).}
\newcommand{\yptrsa}[5]{, ``#5,'' {\em Phil.\ Trans.\ Roy.\ Soc.\ A }{\bf #2}, #3--#4 (#1).}
\newcommand{\yrmp}[5]{, ``#5,'' {\em Rev.\ Mod.\ Phys.\ }{\bf #2}, #3--#4 (#1).}
\newcommand{\yjcp}[5]{, ``#5,'' {\em J.\ Comp.\ Phys.\ }{\bf #2}, #3--#4 (#1).}
\newcommand{\yjgr}[5]{, ``#5,'' {\em J.\ Geophys.\ Res.\ }{\bf #2}, #3--#4 (#1).}
\newcommand{\ygrl}[5]{, ``#5,'' {\em Geophys.\ Res.\ Lett.\ }{\bf #2}, #3--#4 (#1).}
\newcommand{\ygrlN}[4]{, ``#4,'' {\em Geophys.\ Res.\ Lett.\ }{\bf #2}, #3 (#1).}
\newcommand{\yobs}[5]{, ``#5,'' {\em Observatory }{\bf #2}, #3--#4 (#1).}
\newcommand{\yjas}[5]{, ``#5,'' {\em J.\ Atmosph.\ Sci.\ }{\bf #2}, #3--#4 (#1).}
\newcommand{\yaj}[5]{, ``#5,'' {\em Astronom.\ J.\ }{\bf #2}, #3--#4 (#1).}
\newcommand{\sapj}[3]{, ``#3,'' {\em Astrophys.\ J.\ }, submitted, arXiv:#2 (#1).}
\newcommand{\tapj}[2]{, ``#2,'' {\em Astrophys.\ J.\ }, to be submitted (#1).}
\newcommand{\sapjl}[3]{, ``#3,'' {\em Astrophys.\ J.\ Lett.}, submitted, arXiv:#2 (#1).}
\newcommand{\papjl}[3]{, ``#3,'' {\em Astrophys.\ J.\ Lett.}, in press, arXiv:#2 (#1).}
\newcommand{\yapj}[5]{, ``#5,'' {\em Astrophys.\ J.\ }{\bf #2}, #3--#4 (#1).}
\newcommand{\yapjN}[4]{, ``#4,'' {\em Astrophys.\ J.\ }{\bf #2}, #3 (#1).}
\newcommand{\yapjlN}[4]{, ``#4,'' {\em Astrophys.\ J.\ Lett.\ }{\bf #2}, #3 (#1).}
\newcommand{\yapjlNS}[4]{, ``#4'' {\em Astrophys.\ J.\ Lett.\ }{\bf #2}, #3 (#1).}
\newcommand{\yapjS}[5]{, ``#5,'' {\em Astrophys.\ J.\ }{\bf #2}, #3--#4 (#1)}
\newcommand{\yapjlS}[5]{, ``#5,'' {\em Astrophys.\ J.\ Lett.\ }{\bf #2}, #3--#4 (#1)}
\newcommand{\yapjs}[5]{, ``#5,'' {\em Astrophys.\ J.\ Suppl.\ }{\bf #2}, #3--#4 (#1).}
\newcommand{\yapjl}[5]{, ``#5,'' {\em Astrophys.\ J.\ Lett.\ }{\bf #2}, #3--#4 (#1).}
\newcommand{\ypp}[5]{, ``#5,'' {\em Phys.\ Plasmas }{\bf #2}, #3--#4 (#1).}
\newcommand{\ypac}[5]{, ``#5,'' {\em Publ.\ Astron.\ Soc.\ Pacific }{\bf #2}, #3--#4 (#1).}
\newcommand{\yaraa}[5]{, ``#5,'' {\em Ann.\ Rev.\ Astron.\ Astrophys.\ }{\bf #2}, #3--#4 (#1).}
\newcommand{\yarXiv}[3]{:~#1, ``#3,'' arXiv:#2}
\newcommand{\yjcap}[4]{, ``#4,'' {\em J.\ Cosm.\ Astrop.\ Phys.\ }{\bf #2}, #3 (#1).}
\newcommand{\ycqg}[5]{, ``#5,'' {\em Class.\ Quant.\ Grav.\ }{\bf #2}, #3--#4 (#1).}
\newcommand{\yanar}[5]{, ``#5,'' {\em Astron.\ Astrophys.\ Rev.\ }{\bf #2}, #3--#4 (#1).}
\newcommand{\yanf}[5]{, ``#5,'' {\em Ann.\ Rev.\ Fluid Dyn.\ }{\bf #2}, #3--#4 (#1).}
\newcommand{\ypf}[5]{, ``#5,'' {\em Phys.\ Fluids }{\bf #2}, #3--#4 (#1).}
\newcommand{\yija}[5]{, ``#5,'' {\em Int.\ J.\ Astrobiol.\ }{\bf #2}, #3--#4 (#1).}
\newcommand{\yijaS}[5]{, ``#5'' {\em Int.\ J.\ Astrobiol.\ }{\bf #2}, #3--#4 (#1).}
\newcommand{\ypfN}[4]{, ``#4,'' {\em Phys.\ Fluids }{\bf #2}, #3 (#1).}
\newcommand{\yphy}[5]{, ``#5,'' {\em Physica } {\bf #2}, #3--#4 (#1).}
\newcommand{\ygafd}[5]{, ``#5,'' {\em Geophys.\ Astrophys.\ Fluid Dynam. }{\bf #2}, #3--#4 (#1).}
\newcommand{\yoleb}[5]{, ``#5,'' {\em Orig.\ Life Evol.\ Biosph. }{\bf #2}, #3--#4 (#1).}
\newcommand{\yab}[5]{, ``#5,'' {\em Astrobiol. }{\bf #2}, #3--#4 (#1).}
\newcommand{\yzfa}[5]{, ``#5,'' {\em Zeitschr.\ f.\ Astrophys.\ }{\bf #2}, #3--#4 (#1).}
\newcommand{\yptp}[5]{, ``#5,'' {\em Progr.\ Theor.\ Phys.\ }{\bf #2}, #3--#4 (#1).}
\newcommand{\ypfb}[5]{, ``#5,'' {\em Phys.\ Fluids B }{\bf #2}, #3--#4 (#1).}
\newcommand{\yjour}[6]{, ``#6,'' {\em #2} {\bf #3}, #4--#5 (#1).}
\newcommand{\yjourN}[5]{, ``#5,'' {\em #2} {\bf #3}, #4 (#1).}
\newcommand{\yjourNN}[4]{, ``#4,'' {\em #2}, p.~#3 (#1).}
\newcommand{\yjourS}[6]{, ``#6'' {\em #2} {\bf #3}, #4--#5 (#1).}
\newcommand{\pjour}[3]{, ``#3,'' {\em #2}, in press (#1).}
\newcommand{\djour}[4]{, ``#3,'' {\em #2}, DOI:#4 (#1).}
\newcommand{\dnjour}[5]{, ``#4,'' {\em #2}, {\bf #3}. DOI:#5 (#1).}
\newcommand{\ppre}[3]{, ``#2,'' {\em Phys.\ Rev.\ E}, in press, arXiv:#3 (#1).}
\newcommand{\sjour}[3]{, ``#3,'' {\em #2}, submitted (#1).}
\newcommand{\yprep}[2]{, ``#2,'' (#1) (preprint).}
\newcommand{\pproc}[5]{, ``#2,'' In {\em #3} (ed.\ #4), #5 (#1) (to appear).}
\newcommand{\yproc}[7]{, ``#4,'' In {\em #5} (ed.\ #6), pp.\ #2--#3.\ #7 (#1).}
\newcommand{\ybook}[3]{ {\em #2}.\ #3 (#1).}
\def\blue{\textcolor{blue}}
\def\apjs{ApJS}
\def\pre{PhRvE}


\title{\MakeLowercase{\mbox{$f$}}-mode strengthening from a localised bipolar subsurface magnetic field}

\author{Nishant K. Singh$^{1 \ast}$,\thanks{$^\ast$Corresponding author.
Email: singh@mps.mpg.de\vspace{6pt}}
Harsha Raichur$^1$,
Maarit J. K\"apyl\"a$^{1,2}$,
Matthias Rheinhardt$^2$,
Axel~Brandenburg$^{3,4,5,6}$ 
and Petri J. K\"apyl\"a$^{7,2}$ \\ \vspace{6pt}
$^1$Max Planck Institute for Solar System Research,
              37077 G\"ottingen, Germany\\
$^2$ReSoLVE Centre of Excellence, Dept. of Computer Science,
              Aalto University, FI-00076, Finland\\
$^3$NORDITA, KTH Royal Inst. of Technology and Stockholm University,
              10691 Stockholm, Sweden\\
$^4$Dept. of Astronomy, AlbaNova University Center,
              Stockholm University, 10691 Stockholm, Sweden\\
$^5$JILA and Dept. of Astrophysical and Planetary Sciences,
              University of Colorado, Boulder, USA\\
$^6$Laboratory for Atmospheric and Space Physics,
              3665 Discovery Drive, Boulder, CO 80303, USA\\
$^7$Georg-August-Universit\"at G\"ottingen, Institut f\"ur
              Astrophysik, Friedrich-Hund-Platz 1, D 37077 G\"ottingen,
              Germany\\
  \vspace{6pt}\received{\it \today,~ $ $Revision: 1.203 $ $} }

\maketitle

\begin{abstract}
Recent numerical work in helioseismology
has shown that a periodically varying subsurface magnetic
field leads to a fanning of the $f$-mode, which emerges from a density
jump at the surface.
In an attempt to model a more realistic situation, we now modulate this
periodic variation with an envelope, giving thus more emphasis on
localised bipolar magnetic structures in the middle of the domain.
Some notable findings are: (i) compared to the purely hydrodynamic
case, the strength of the $f$-mode is significantly larger at high
horizontal wavenumbers $k$, but the fanning is weaker for the localised
subsurface magnetic field concentrations investigated here than the
periodic ones studied earlier;
(ii) when the strength of the
magnetic field is enhanced at a fixed depth below the surface, the
fanning of the $f$-mode in the $k\omega$ diagram
increases proportionally in such a way that the 
normalised
$f$-mode strengths
remain nearly the same in different such cases;
(iii) the unstable Bloch modes reported previously in case of harmonically
varying magnetic fields are now completely absent when more realistic
localised magnetic field concentrations are imposed beneath the surface,
thus suggesting that the Bloch modes are unlikely to be supported
during most phases of the solar cycle;
(iv) the $f$-mode strength appears to depend also on the depth of
magnetic field concentrations such that it shows a relative decrement when the
maximum of the magnetic field is moved to a deeper layer.
We argue that detections of $f$-mode perturbations such as those
being explored here could be effective tracers of solar magnetic
fields below the photosphere before these are directly detectable
as visible manifestations in terms of active regions or sunspots.
\end{abstract}


\section{Introduction}

The Sun supports a wide variety of waves that carry useful
information about the internal solar structure, which can be
inferred by employing the methods of helioseismology
\citep[see, e.g.,][]{G87,C03}.
Local analysis using especially the surface gravity  
mode (or the $f$-mode)
is useful in studying the near-surface structure
\citep[e.g.][]{HBBG08,FBCB12,FCB13,DABCG11}.
Some properties of surface waves in idealised settings involving
magnetic fields were explored in \citet{C61,R81,MR89,MR92,MAR92}.
It is of great interest to probe interior magnetic fields of
the Sun using the techniques of helioseismology; see \cite{T06} for
a review on the subject of magnetohelioseismology.
Systematic changes, notably in the frequencies of the global
$f$-mode, as a function of solar cycle
were found and discussed in detail in \citet{T06} and \citet{P08}.
\cite{T06} found evidence of a $500\G$ magnetic field at a depth
of about $5\Mm$ and suggested $2\%$ modulation in turbulent
convection velocities from solar minimum to maximum,
based on the observed cycle-dependence
of mean frequency shifts of the $f$- and $p$-modes.
Moreover, it is reasonable to expect magnetically induced variations
in the $f$-mode on much shorter timescales, e.g.,
during localised magnetic flux emergence leading to the
formation of active regions (ARs) or sunspots.

Much of the earlier studies on the global $f$-mode of the Sun
focussed primarily on the frequency shifts that were observed
\citep{LWK90,FSTT92}. The frequencies were significantly smaller
than the theoretically expected values, where both the shift and line width
grow with the spherical harmonic degree. Subsequent studies explained
these findings by invoking turbulent background motions and deriving
a generalised dispersion relation of the $f$-mode in the presence of
a random velocity field \citep{MR93a,MR93b,MMR99,M00a,M00b,MKE08}.
The influence of coherent (for modelling supergranulation)
as well as random (mimicking near-surface granulation) flows
on the $f$-mode were explored in detail by \cite{M00a} where
it was found that, while a space-dependent random flow causes
a decrement, a time-dependent random flow can enhance the
frequencies. Observations were thus explained in terms of
the parameters of the chosen velocity field.
However, these studies ignored the effect of magnetic fields, which
can increase the $f$-mode frequencies \citep[e.g.][]{C61,R81,MR92,MAR92}.

In a series of works, \citet{CB93,CB97} and \citet{CBZ94}
investigated the interaction
of $f$- and $p$-modes with a vertical magnetic field and found that
a partial conversion of these modes into slow magnetoacoustic modes
takes place whenever they encounter a vertical field resembling
those of sunspots. \cite{PK09} numerically explored the effects
of inclined magnetic fields and noted that the $f$-modes are more 
strongly affected by the background magnetic field than the $p$-modes.
\cite{SBCR15} studied numerically various properties of the $f$-
as well as $p$- and $g$-modes in a wide variety of magnetic backgrounds.
They found that horizontal magnetic fields cause an increase in
the $f$-mode frequencies -- as expected. But their dependencies are
more complicated in the presence of vertical or oblique magnetic fields,
which may be more relevant for the predominantly vertical
fields of sunspots.
In this case, the $f$-mode frequencies are enhanced
relative to their nonmagnetic values
at intermediate horizontal wavenumbers, but decreased 
at large wavenumbers.

In the presence of magnetic fields in the solar atmosphere,
Alfv\'en and magnetosonic waves are known to couple resonantly
with the global oscillations, affecting mainly the frequencies,
line widths, and penetration of the $f$ and $p$ modes into the
solar atmosphere
\citep[see][for a review and references therein]{E06}.
Another important finding was that 
resonant interaction between
global modes and 
Alfv\'en waves
causes a damping of the $f$- and $p$-modes due to dissipative effects near
the resonance frequency.
\cite{PE18} further study this situation 
by deriving the dispersion relation of the  $f$-mode
in a magnetically coupled solar interior-atmosphere system to
obtain its frequency shifts. For a magnetised atmosphere,
these were found to be positive relative to an unmagnetised one.

It is of great interest to use numerical simulations to study 
the effects of subsurface magnetic fields on 
both the acoustic or $p$-modes and the $f$-mode.
Such studies aim
at refining the interpretation of helioseismic measurements
which use
sound waves to infer the internal structure of the Sun
and its internal motions \citep[e.g.][]{Ba16,HGR16}.
This is necessary, because magnetic fields
complicate the usage of helioseismic inversion techniques, as their 
presence gives rise to 
the modification of sound and gravity waves into
magnetoacoustic and magnetogravity 
ones
which are difficult to account for \citep{T83,C11}.
Furthermore, if the properties and behaviour of $p$- and $f$-modes
in the presence of magnetic fields of different strength, topology 
or location, in particular depth,
were known well enough from the
numerical models, we might be able to infer the subsurface magnetic 
fields from  helioseismic measurements. From such inferences, we
may be able to learn about the origin of subsurface magnetic fields,
that is, about the solar dynamo mechanism \citep{B05,Cha10,KMB12}, and about the
process of concentrating magnetic fields into sunspots and
ARs; see, e.g., \cite{BRK16} and \cite{KBKKR16} for
a competing
mechanism enabling a self-consistent formation of localised magnetic flux
concentrations.

Such numerical simulations, enabling us to study the effects of magnetic
fields on the naturally occurring modes of oscillations,
have recently been performed with the
{\sc Pencil Code}\footnote{http://github.com/pencil-code}.
\cite{SBCR15} introduced a modelling framework with a piecewise isothermal
atmosphere, where the upper layer is mimicking a hot
corona and the lower one a (cooler) convection zone.
The two layers are separated by a jump in density and
temperature, which represents the solar surface and enables the
presence of the  $f$-mode. For a simple
approximation of the convective turbulence, 
random hydrodynamic forcing was applied in the lower layer.
In such a setup, acoustic ($p$), internal gravity ($g$), and
surface gravity ($f$) modes are all {\it self-consistently} driven, 
in contrast to
other types of approaches, which selectively produce the modes based on
linearised equations \citep{DABCG11,SCGM11}.
\cite{SBCR15} studied the influence of uniformly imposed magnetic fields on
the $p$- and $f$-mode properties, and verified the
expectation of $f$-modes being
sensitive to the presence of magnetic fields and $p$-modes 
being less affected.
The major effect on the $f$-modes was an increase in
their mode frequencies. 
\cite{SBR14} developed the setup further to include non-uniform
magnetic fields with
harmonic profiles within the convection zone. This work revealed
the {\it fanning effect} of the $f$-mode, that is, an increase of its line width
towards higher wavenumbers. 
The width of the fan and its asymmetry could be directly related to the
strength and location of the magnetic field.

These numerical studies and their predictions led to an
observational case study of $f$-modes in relation to the emergence
of about half a dozen ARs, observed with
HMI, and the resulting line-of-sight Dopplergrams and
magnetograms \citep{SRB16}.
This study reported
\emph{strengthening} of the local $f$-mode
about two days before the emergence of ARs at the same
corotating patch. It was noted that such a precursor signal can be
detected by isolating high-degree $f$-modes through careful fitting and
subsequent subtraction of background and $p$-modes.
It was argued there that the precursor signal is best seen when an AR
forms in isolation, i.e., far from other existing ARs which can `pollute'
the signal. It is known that the sunspots or ARs absorb the $f$-mode
power, thus causing its damping \citep{CB97}, and therefore it is expected
to be harder to extract the precursor signal associated with a
newly forming AR in a `crowded' environment with many existing ARs.
A possible cleaning procedure to still extract the signal was discussed
in \cite{SRB16}.
Although this is yet to be confirmed
for larger data sets and with independent observational techniques,
the potential significance
of such $f$-mode related precursors
cannot be understated.
Based on the photospheric velocity measurements, \cite{KT17}
detected plasma upflows somewhat before the emergence of two small
ARs.
A number of previous case studies have reported detections of
subphotospheric velocities associated with emerging ARs
using different techniques, such as,
helioseismic holography, ring-diagram or time-distance analysis
\citep{Komm08,Har11,Ilo11,Birch13,Barnes14}.
Given that the $f$-mode eigenfunctions extend to depths of about
a few Mm, it might be expected to be sensitive to
such velocity perturbations.

In this paper we extend the model of \cite{SBR14,SBCR15} to
study the $f$-mode strengthening
and fanning using inhomogeneous magnetic fields of different
strength, topology, and location in the convection zone.
In addition to using harmonic profiles which stretch over
the whole horizontal extent, we use localised harmonic perturbations,
to better mimic isolated active regions.
We describe our model and basic definitions in \Sec{model} and present
our results in \Sec{res}.
Conclusions are given in \Sec{conc}.

\section{Model setup and analysis technique}
\label{model}

We consider here a model that is similar to that studied in
\cite{SBR14,SBCR15}. The $p$-, $g$- and $f$-modes are produced in
a self-consistent manner in a two-dimensional Cartesian $x$-$z$ domain
with a piecewise isothermal medium consisting of a cool lower layer,
called bulk, and a hotter upper layer, called corona.
The thicknesses of these layers are $\Ld$ and $L_{\rm u}$,
respectively, where subscripts `d' and `u' refer to the 
layers downward and upward of the interface.
A non-uniform magnetic background field is maintained
by including an electromotive force (EMF) in the uncurled version
of the induction equation for the magnetic vector potential $\bm A$.
We solve the basic hydromagnetic equations,
\begin{align}
\frac{D \ln \rho}{Dt} &= -\bm\nabla\cdot\bm{u}, \\
\frac{D\bm{u}}{Dt} &= \ff+\bm{g} +\frac{1}{\rho}
\left(\bm{J}\times\bm{B}-\bm\nabla p
+\bm\nabla \cdot 2\nu\rho\bm{\mathsf{S}}\right), \label{mom-eq}\\
T\frac{D s}{Dt} &= 2\nu \bm{\mathsf{S}}^2 + \frac{\mu_0\eta}{\rho}\JJ^2
- (\gamma-1) c_p \frac{T-T_{\rm d,u}}{\tau_{\rm c}}\, ,
\label{equ:ss} \\
\frac{\partial \bm A}{\partial t} &= {\bm u}\times{\bm B} 
+ \EMF_0- \eta \mu_0 {\bm J}, \label{ind_eq}
\end{align}
where $\rho$ is the density, $\bm{u}$ is the velocity,
$D/Dt = \partial/\partial t + \bm{u} \cdot \bm\nabla$ is the
advective time derivative,
$\bm{f}$ is a forcing function [as specified in \citet{B01}] to drive
seismic modes of low Mach number\footnote{Note that here
the vector $\ff$ denotes the forcing in the velocity equation,
whereas the scalar $f$ refers to the surface gravity $f$-mode.},
$\bm{g}=(0,0,-g)$ is the gravitational acceleration,
$\mathsf{S}_{ij}=\half(u_{i,j}+u_{j,i})
-\onethird \delta_{ij}\bm\nabla\cdot\bm{u}$
is the traceless rate of strain tensor, with commas denoting
partial differentiation,
$\nu=\const$ is the kinematic viscosity,
${\bm B} =\bm\nabla\times{\bm A}$ is the magnetic field,
${\bm J} =\mu_0^{-1}\bm\nabla\times{\bm B}$ is the current density,
$\mu_0$ is the vacuum permeability,
$s$ is the specific entropy,
$\gamma=\cp/\cv$ is the ratio of specific heats
at constant pressure and volume, respectively,
 $T$ is the temperature,
$\EMF_0$ is an external EMF specified below,
and $\eta=\const$ is the magnetic diffusivity.
Since the forcing amplitude is small, no additional hyperdiffusion is
needed, and also no diffusion in the energy equation was included.
The fluid is assumed to obey the equation of state of an ideal gas, hence
the pressure is given by $p=(\cp-\cv)\rho T=\rho \cs^2/\gamma$.
The medium is vertically stratified under constant gravity, $g>0$,
where we identify $x$ and $z$ as horizontal and vertical directions,
respectively.
All calculations are performed with the {\sc Pencil Code}.

\begin{figure}
\begin{center}
\includegraphics[width=0.6\textwidth]{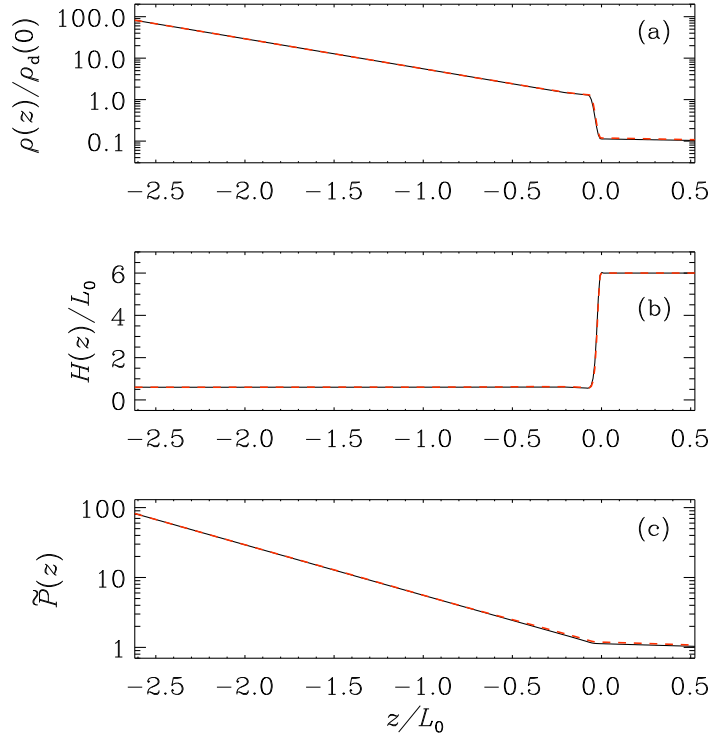}
\caption{Density (a), scale height (b), and pressure (c)
of the horizontally averaged background state as functions of $z$,
where $\tilde{P}=\gamma P/\rho_{\rm d}(0)\csd^2$.
Solid, black: hydro (H1); dashed, red: magnetic (A1) case.
}\label{ther_var}
\end{center}
\end{figure}

The variations of background density, pressure scale height, and
pressure as functions of $z$ are shown in \Fig{ther_var}.
Similar to earlier works \citep{SBR14,SBCR15} 
we introduce a sharp jump in density at the interface $z=0$ with
$\rho_{\rm u}(0)\ll\rho_{\rm d}(0)$, along with corresponding jumps
in temperature. The adiabatic sound speed $\cs$
is maintained by the last term in \Eq{equ:ss}, which
guarantees the relaxation to constant 
average temperatures $T_{\rm d}$ and $T_{\rm u}$ in either subdomain
within a relaxation time $\tau_{\rm c}$
(constant throughout the domain). In this way, the interface is
created and maintained and the $f$-mode is naturally enabled.
It is therefore also
known as the free surface mode.
As the density decreases exponentially with height $z$ in 
an isothermally stratified medium, it is given by
$\rho_{\rm d,u}(z)=\rho_{\rm d,u}(0)\exp (-z/H_{\rm d,u})$, where
$H_{\rm d,u}=(\cp-\cv) T_{\rm d,u}/g$ is the scale height.
(The pressure and density scale heights are equal for an isothermal layer.)
The sharp jump in the thermodynamic quantities is quantified by the
ratio
\EQ
q=\frac{\rho_{\rm u}(0)}{\rho_{\rm d}(0)}=\frac{\csd^2}{\csu^2}
=\frac{T_{\rm d}}{T_{\rm u}}=\frac{H_{\rm d}}{H_{\rm u}}.
\label{q}
\EN
We enforce a steady magnetic field $\BB_0$ by applying a constant external
EMF $\EMF_0=\emf_{0y}^{\rm A,B} \ee_y$.
Two different types of magnetic structures are considered
in the magnetohydrodynamic (MHD) simulations, called models~A and B.
Model A is characterized by a harmonic variation in both spatial dimensions
\EQ
\emfA=\hat{\cal E}_0 \cos \left(k^B_x x\right) \cos \left(k^B_z z\right),
\label{emfA}
\EN
and model B by a localised sine wave
\EQ
\emfB=\hat{\cal E}_0 \FTH \sin \left(k^B_x x\right)
\exp{\left[-\left(\frac{z-z_*}{w_z}\right)^2\right]}\,,
\label{emfB}
\EN
where $\FTH$ is a smoothed top hat as a function of $x$,
centred w.r.t. $x$.
 It is unity for
$x\in[x_1,x_2]$ and smoothly goes to zero outside this interval;
see, e.g., \Figs{snapshots_A}{fmm_BI} for the
profiles of the imposed magnetic
background in its saturated state.
The transitions at $x_1$ and $x_2$ are modelled with a third-order
polynomial of width $w\approx0.3\Hd$.
Note that the sustained magnetic field $\BB_0$
drives a large-scale
flow which, in turn, acts back on the field, so that its shape
is not simply determined by
$\eta\nabla^2\BB_0 + \nabla\times\EMF_0=\boldsymbol{0}$.
The A models are very similar to those considered in
\cite{SBR14}, whereas the B ones were tailored to model
an emerging active region more realistically:
given that ARs hardly ever show a periodic pattern in
longitude, we restricted the horizontal extent
of the subsurface magnetic structures in the B models, 
mimicking more closely the fields of bipolar regions.
Both models allow us to explore the effects of 
subsurface magnetic fields on the
$f$-mode,
naturally occurring at the interface
in our minimalistic setup.
It is nevertheless sufficiently realistic
for understanding the solar $f$-mode that is expected to be unaffected
by the choice of the thermodynamic background state of the medium.
The top and bottom boundaries were chosen to be
stress free and perfectly conducting,
whereas periodicity was assumed in the horizontal direction.

The length scales and frequencies are normalised by
$L_0=\gamma H_{\rm d}=\csd^2/g$ and $\omega_0=g/\csd$, and the
dimensionless variables are indicated by tildae,
i.e., $\widetilde{k}_x=k_x L_0$, $\widetilde{\omega}=\omega/\omega_0$,
and so on. From the vertical velocity $u_z$ at $z=0$, we construct the
diagnostic $k_x$-$\omega$ diagram ($k\omega$ diagram for short)
by taking the Fourier transform of $u_z(x,0,t)$,
giving $\hat{u}_z(k_x,\omega)$, which here has the same dimension as 
velocity.
The $k\omega$ diagrams, such as those shown in \Fig{hydro}
are constructed from the dimensionless quantity
\EQ
\widetilde{P}(k_x,\omega)=\frac{|\hat{u}_z(k_x,\omega)|}{u_0}\,,
\label{Pdef}
\EN
where the mass-weighted root-mean-squared (rms)
velocity $u_0=\sqrt{\langle\rho^2 u_{\rm d}^2\rangle/\langle\rho^2\rangle}$
is employed for normalisation 
and the angle brackets denote volume averaging over the bulk.
We define the fluid Reynolds and Mach numbers
as
$\Rey=u_0/(\nu \kf)$ and $\Ma=u_0/\csd$, respectively, where
we choose $\widetilde{k}_{\rm f}=20$ for the wavenumber of the hydrodynamic
low amplitude nonhelical forcing in \Eq{mom-eq}.

\begin{table}\caption{
Summary of all simulations.
$\tilde{x}$--$\tilde{z}$
domain: $8\pi\times\pi$; grid: $1024\times320$;
$\Prm=1$; $q=0.11$; $|\ff|/g=10^{-4}$ for all runs.
$\widetilde{k}_x^B=0.5$, $\widetilde{k}_z^B=2$ and
$\widetilde{w}_z=0.28$ were chosen in \Eqs{emfA}{emfB};
$\Delta\tilde{x}_B=(x_2-x_1)/L_0$.
}
\vspace{1mm}
\small
\centering
\label{table_runs}
\begin{tabular}{c c c c c c c c c c}
\hline\hline\\[-2mm]
Run & $\tilde{\nu}$ & extent of ${\bm{f}}$ & $\Rey$ & $\Ma$ 
& \multicolumn{4}{c}{magnetic background}& $\Sigma_{\rm f}$\\
  &  &  &  &  &  geometry & $\Delta\tilde{x}_B$ & $\max(\vA/\cs)$ & at depth \\[1mm]
\hline\\[-3mm]
H1 & $2\times10^{-4}$ & full                       & 0.22 & 0.00087    & -- & -- & --                                    & --              & 0.0025 \\
H2 & $4\times10^{-4}$ & full                       & 0.08 & 0.00066    & -- & -- & --                                    & --              & 0.0022 \\
H3 & $4\times10^{-4}$ & $\tilde{z}<0$      & 0.08 & 0.00065    & -- & -- & --                                    & --              & 0.0012 \\
H4 & $4\times10^{-4}$ & $\tilde{z}<-0.2$ & 0.08 & 0.00063    & -- & -- & --                                    & --              & 0.0008 \\
\hline\\[-3mm]
A1 & $2\times10^{-4}$ & full                        & 0.56 & 0.0022      & Eq.~(\ref{emfA})& -- &    0.15     & $1.3\Hd$ & 0.0042\\
A2 & $2\times10^{-4}$ & full                        & 0.60 & 0.0024      & Eq.~(\ref{emfA})& -- &    0.18     & $1.3\Hd$ & 0.0038\\
A3 & $2\times10^{-4}$ & full                        & 0.64 & 0.0026      & Eq.~(\ref{emfA})& -- &    0.24     & $1.3\Hd$ & 0.0036\\
A4 & $2\times10^{-4}$ & full                        & 0.77 & 0.0031      & Eq.~(\ref{emfA})& -- &    0.29     & $1.3\Hd$ & 0.0040\\[0mm]
\hline\\[-3mm]
BI1 & $4\times10^{-4}$ & $\tilde{z}<0$     & 0.10  & 0.0008     & Eq.~(\ref{emfB})& $4\pi$ & 0.12& $2.4\Hd$ & 0.0018\\    
BI2 & $4\times10^{-4}$ & $\tilde{z}<0$     & 0.18  & 0.0014     & Eq.~(\ref{emfB})& $4\pi$ & 0.31& $1.8\Hd$ & 0.0046\\    
BII & $4\times10^{-4}$ & $\tilde{z}<-0.2$  & 0.09  & 0.0007     & Eq.~(\ref{emfB})& $7\pi$ & 0.12& $2.4\Hd$ & 0.0010\\    
\hline\hline
\end{tabular}
\end{table}

In order to characterise the strength of the $f$-mode, we determine the
normalised mode strength $\mu_{\rm f}$ as a function of $\widetilde{k}_x$
by fitting a Lorentzian to the line profile of the $f$-mode along
the frequency axis and subtracting the continuum.
Here we define the mode strength as
\EQ
\mu_{\rm f}(\widetilde{k}_x)= \int \Delta\widetilde{P}(\widetilde{k}_x,\widetilde{\omega})\,
\dd \widetilde{\omega}\,,
\label{muf}
\EN
where $\Delta\widetilde{P}$ denotes the normalised excess amplitude of the
$f$-mode over the continuum at $\widetilde{k}_x$.
This yields a wavenumber-dependent
mode strength in a manner similar to that used in \cite{SRB16} where,
however, a
more involved fitting algorithm was used to isolate the $f$-mode, which
lies much closer to the $p$-modes in a more realistic setting with
an isentropic stratification.
It is useful to define an integrated normalised mode amplitude of the
$f$-mode as,
\EQ
\Sigma_{\rm f}= \int \mu_{\rm f} \,\dd \widetilde{k}_x,
\label{sigf}
\EN
which is called here, for short, the {\it relative mode amplitude}.

\section{Results}
\label{res}

We present results from a suite of hydrodynamic (H)
as well as MHD runs (Sets~A and B) with different geometries of
the imposed background
magnetic field. We refer the reader to \Tab{table_runs} for the parameters
of all simulations.

\subsection{Hydrodynamic runs}

\begin{figure}
\begin{center}
\includegraphics[width=0.49\textwidth]{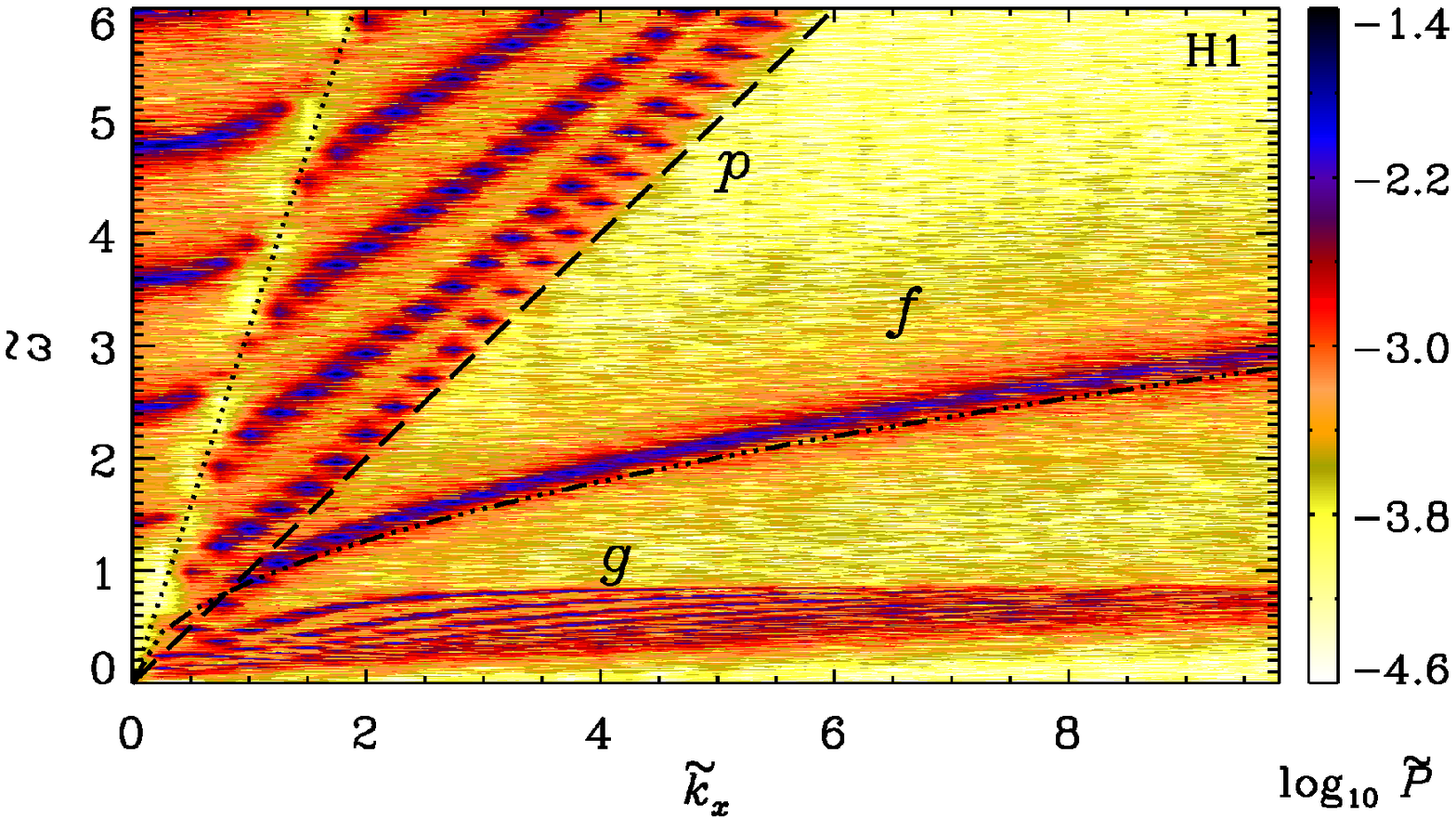}
\includegraphics[width=0.49\textwidth]{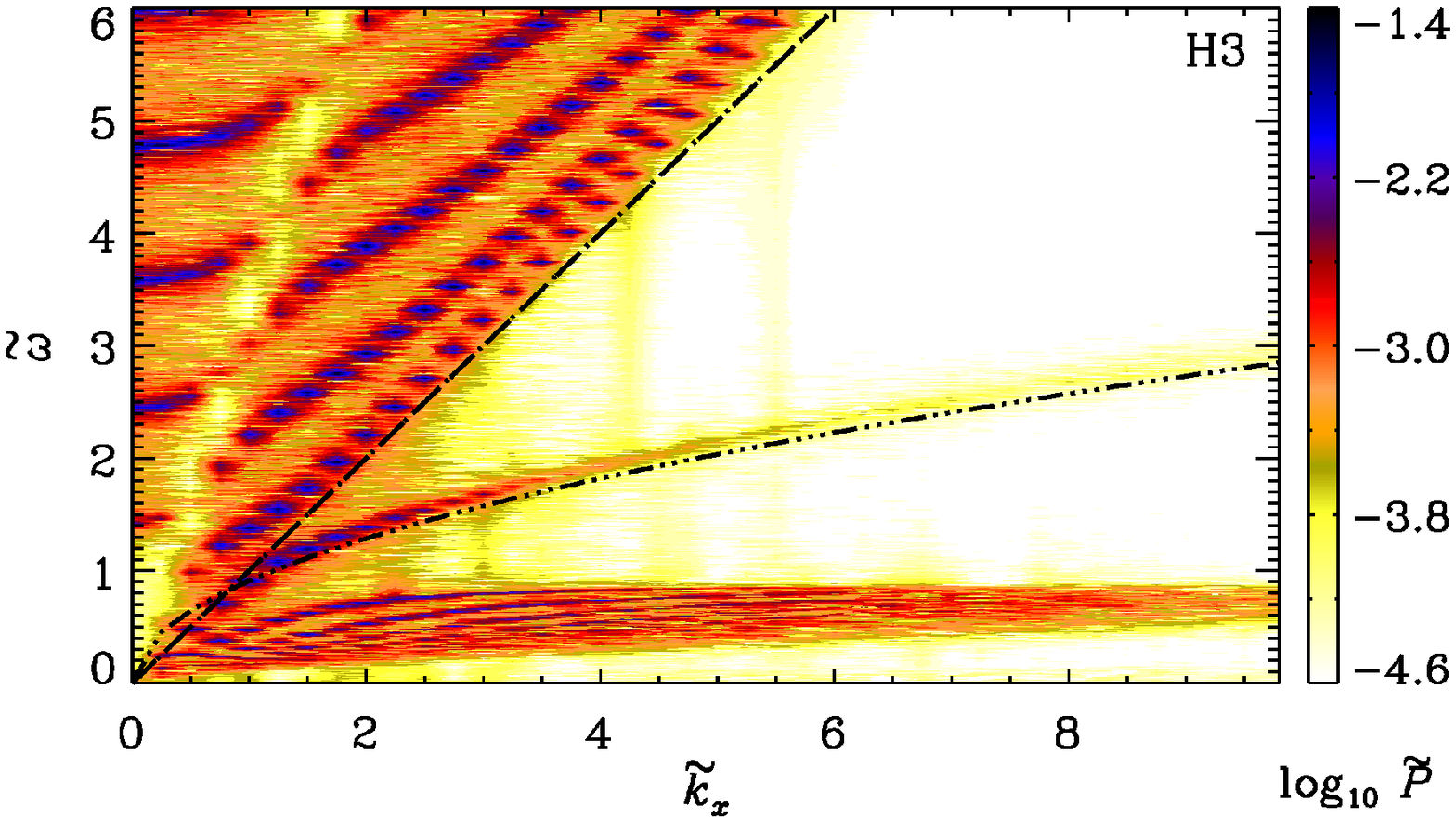}
\vskip0.15in
\includegraphics[width=0.49\textwidth]{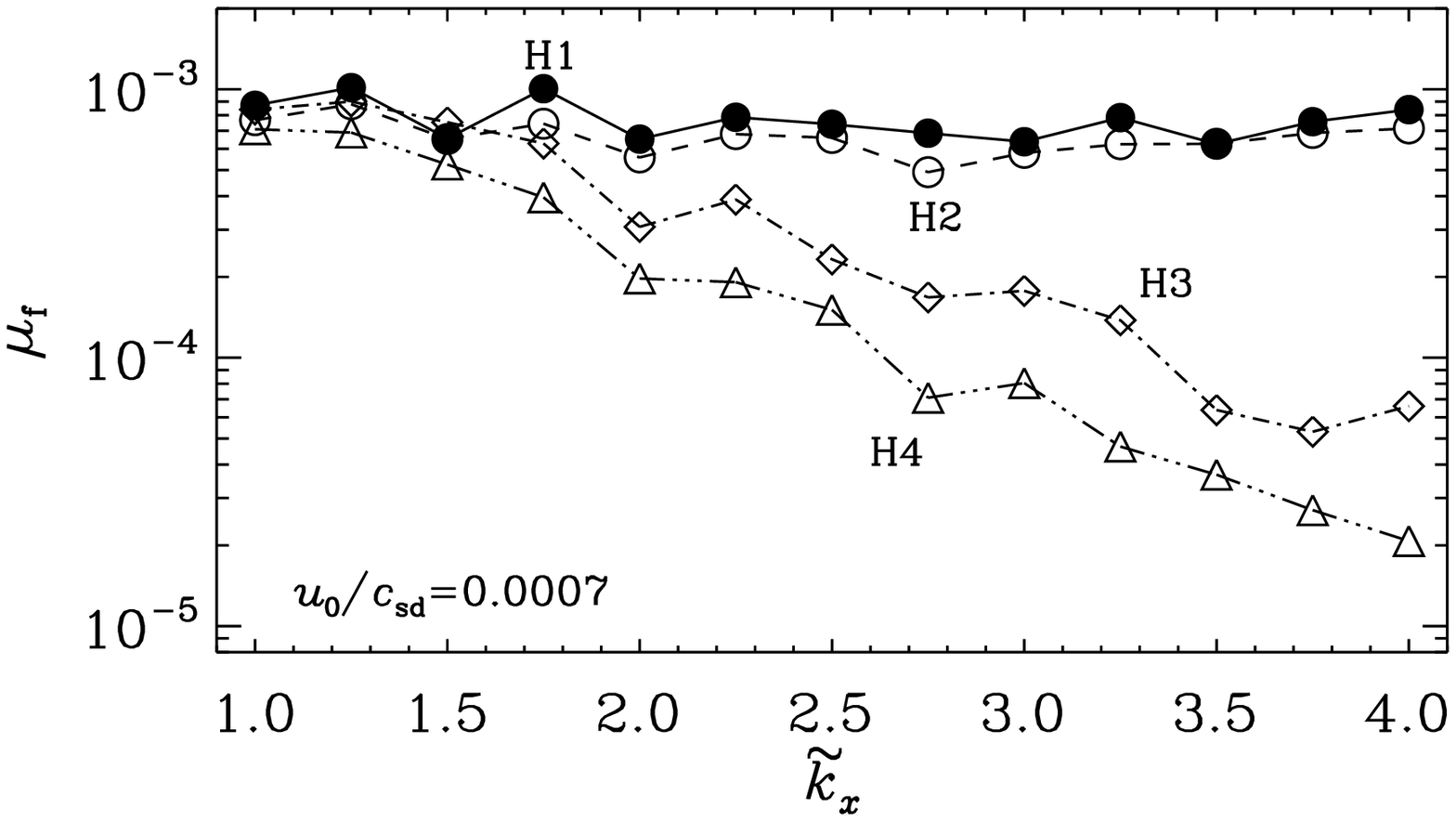}
\includegraphics[width=0.49\textwidth]{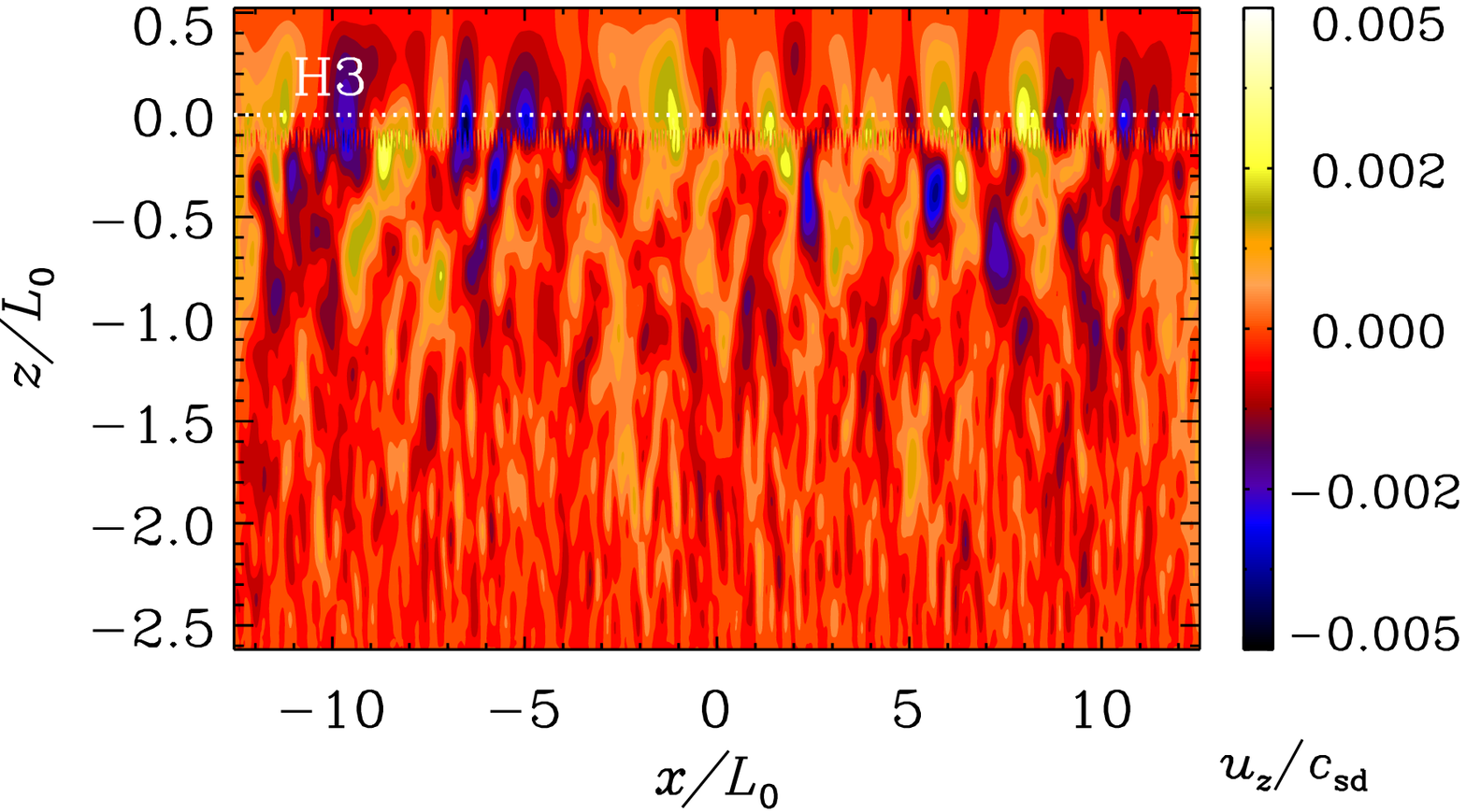}
\caption{Top: $k\omega$ diagrams for the hydrodynamic runs H1 (left) and
H3 (right).
The dotted and dashed lines show $\omega=\csu k_x$ and $\omega=\csd k_x$,
respectively; the dash-dotted curves show $\omega_{\rm f}(k_x)$
of the classical $f$-mode.
Bottom: 
normalised
mode strength $\mu_f$ of the $f$-mode as a
function of $\widetilde{k}_x$ for the
hydrodynamic runs H1--H4 (left), and a snapshot of $u_z/\csd$ in the
statistically
steady state of stochastic motions for the run H3 (right);    
see \Tab{table_runs} for more details.
}\label{hydro}
\end{center}
\end{figure}

We performed four different hydrodynamic runs, H1--H4, to assess
the roles of viscosity and vertical extent of $\ff$.
We first show the diagnostic $k\omega$ diagram, which clearly
reveals the $p$, $g$, and $f$-modes; see the top left
panel of \Fig{hydro} for model~H1.
While the $g$-modes are confined to frequencies
$\widetilde{\omega}<1$, the $p$-modes lie above the line
$\widetilde\omega=\widetilde{k}_x$, as already seen in \cite{SBCR15}
and as expected in the piecewise isothermal setup being considered here.
As the present paper is focussed on the $f$-mode and its interaction
with the subsurface magnetic fields, 
we refer the reader to
\cite{SBCR15} for more details on the properties of the $p$ and $g$-modes.
The $f$-mode in the hydrodynamic cases (referred to as the {\it classical}
$f$-mode) appears close to the theoretical
curve given by 
\EQ
\omega_{\rm f}^2 = g k_x \, \frac{1-q}{1+q}\,,
\label{dr-f}
\EN
\citep[see, e.g.,][]{C61,G87}.

We determine the strength of the $f$-mode using the $k\omega$
diagram as discussed in the previous section, and show the wavenumber
dependence of its 
normalised
mode strength $\mu_{\rm f}$
in the bottom left panel of \Fig{hydro}.
No error bars have been added in this and similar figures below,
because the actual fitting errors are small. 
However, in view of
the non-smooth nature of the curves,
surely other imperfections are present like, e.g., 
the limited integration time of the runs.
The mode strengths from models H1 and H2 are nearly the same;
the viscosity of H2 is twice as large as that of H1.
This is also reflected in the
relative mode amplitude $\Sigma_{\rm f}$ listed in
\Tab{table_runs}. In both these cases, the same hydrodynamic forcing
was employed in the whole domain.
Unlike in models H1 and H2, the $f$-mode strength decreases systematically
with wavenumber $\widetilde{k}_x$ when the forcing is restricted
to layers below the interface, 
namely to $\tilde{z}<0$ in model H3 but to $\tilde{z}<-0.2$ in model H4.
Compared to H3, $\mu_{\rm f}$ is smaller in H4 at nearly all the wavenumbers.
In \Fig{hydro} we also show a snapshot of the vertical motions
in a representative model (H3).
This figure demonstrates that no large scale flow patterns
emerge in the HD simulations.

We note that the mode damping can possibly occur due to a resonant
coupling of atmospheric Alfv\'en and magnetosonic waves
with the global oscillations
of the Sun, in which case dissipative effects play a major role;
see \cite{PEG07}. This is not applicable to the present work as we
study the effects of magnetic fields that are confined to regions below the
surface with negligible leakage into the atmosphere above.
Such cases involving magnetic fields are presented in the following
subsections below.

\subsection{MHD A-type runs}

\begin{figure}
\begin{center}
\includegraphics[width=0.49\textwidth]{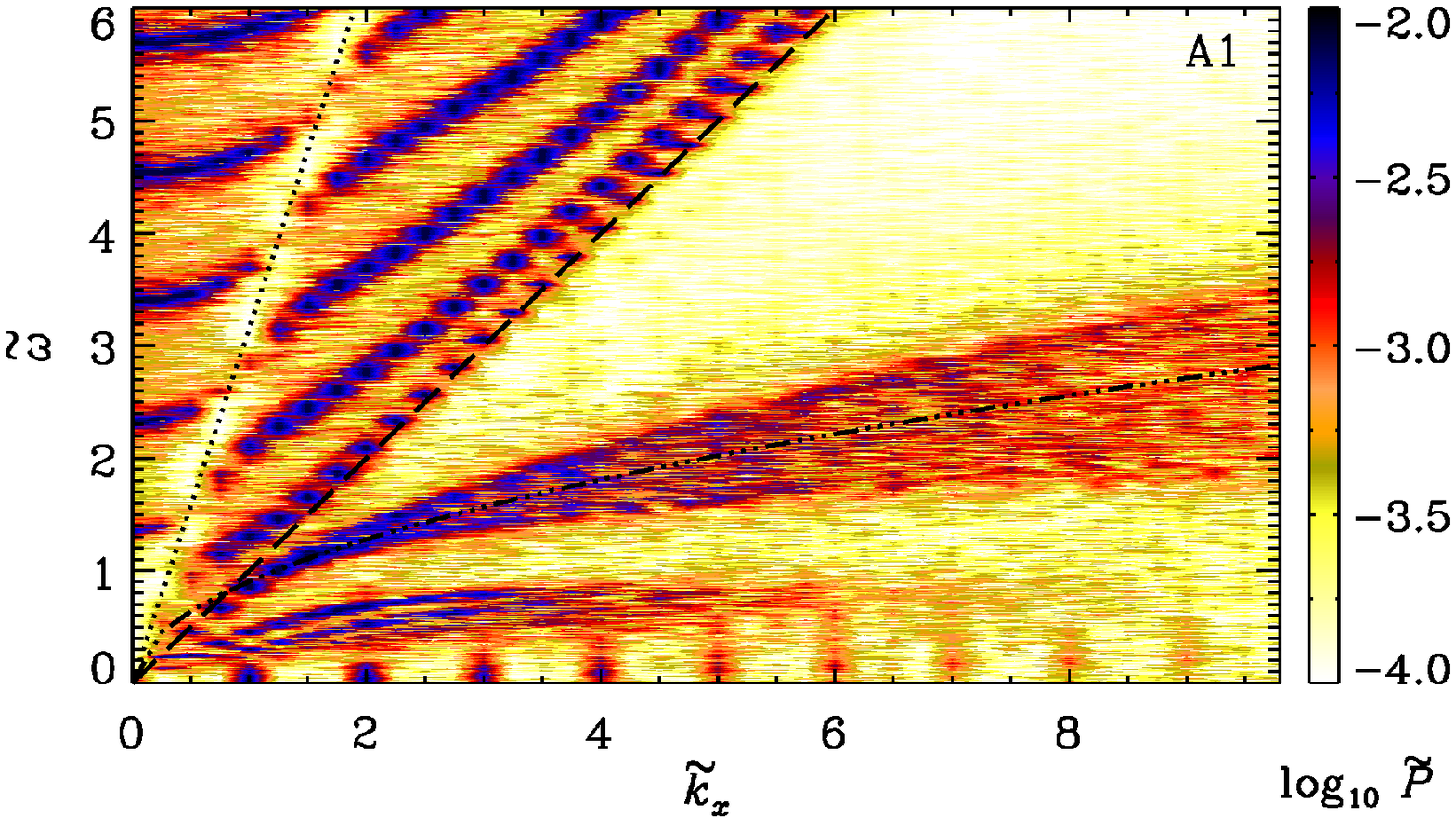}
\includegraphics[width=0.49\textwidth]{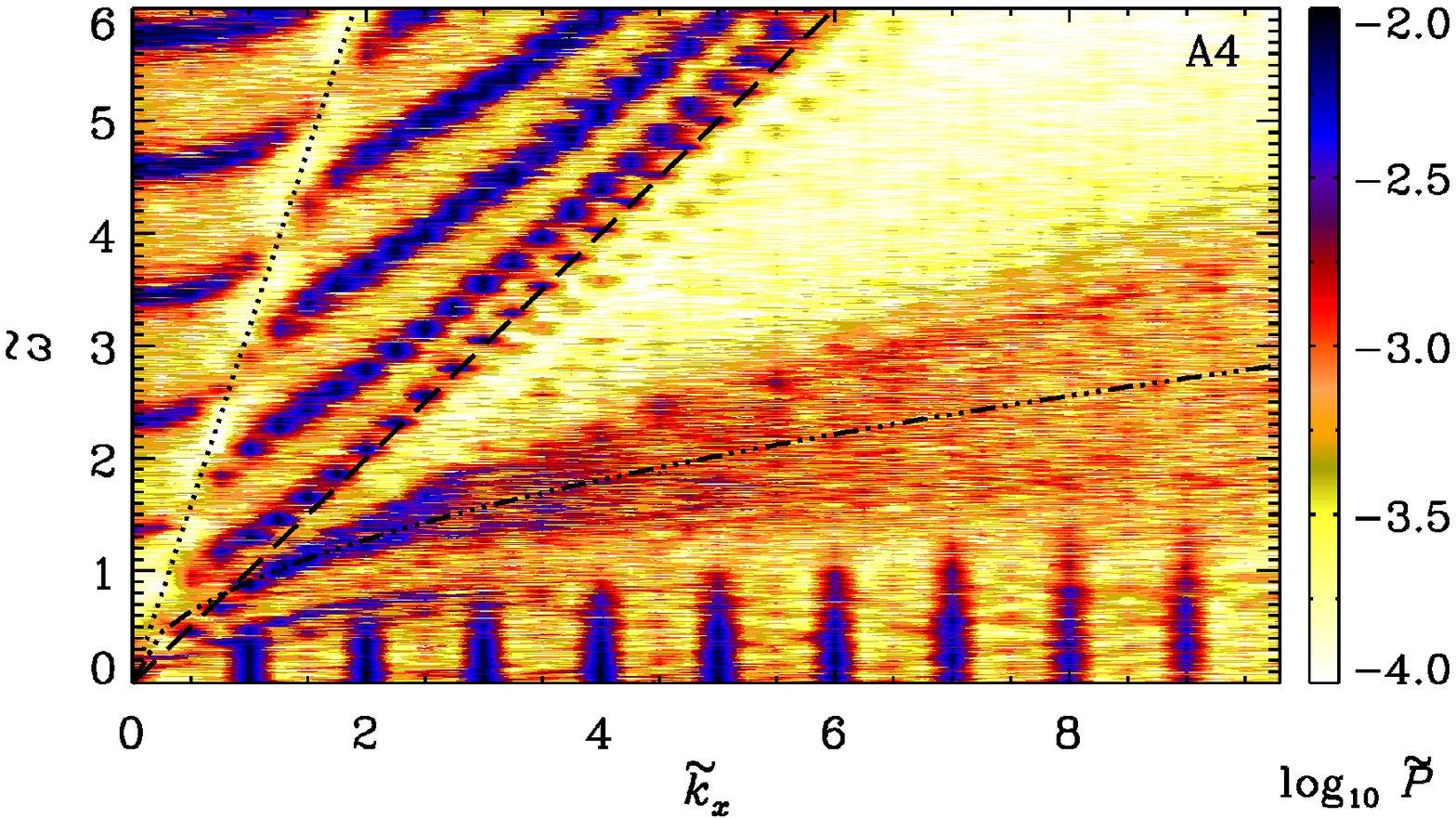}
\vskip0.15in
\includegraphics[width=0.49\textwidth]{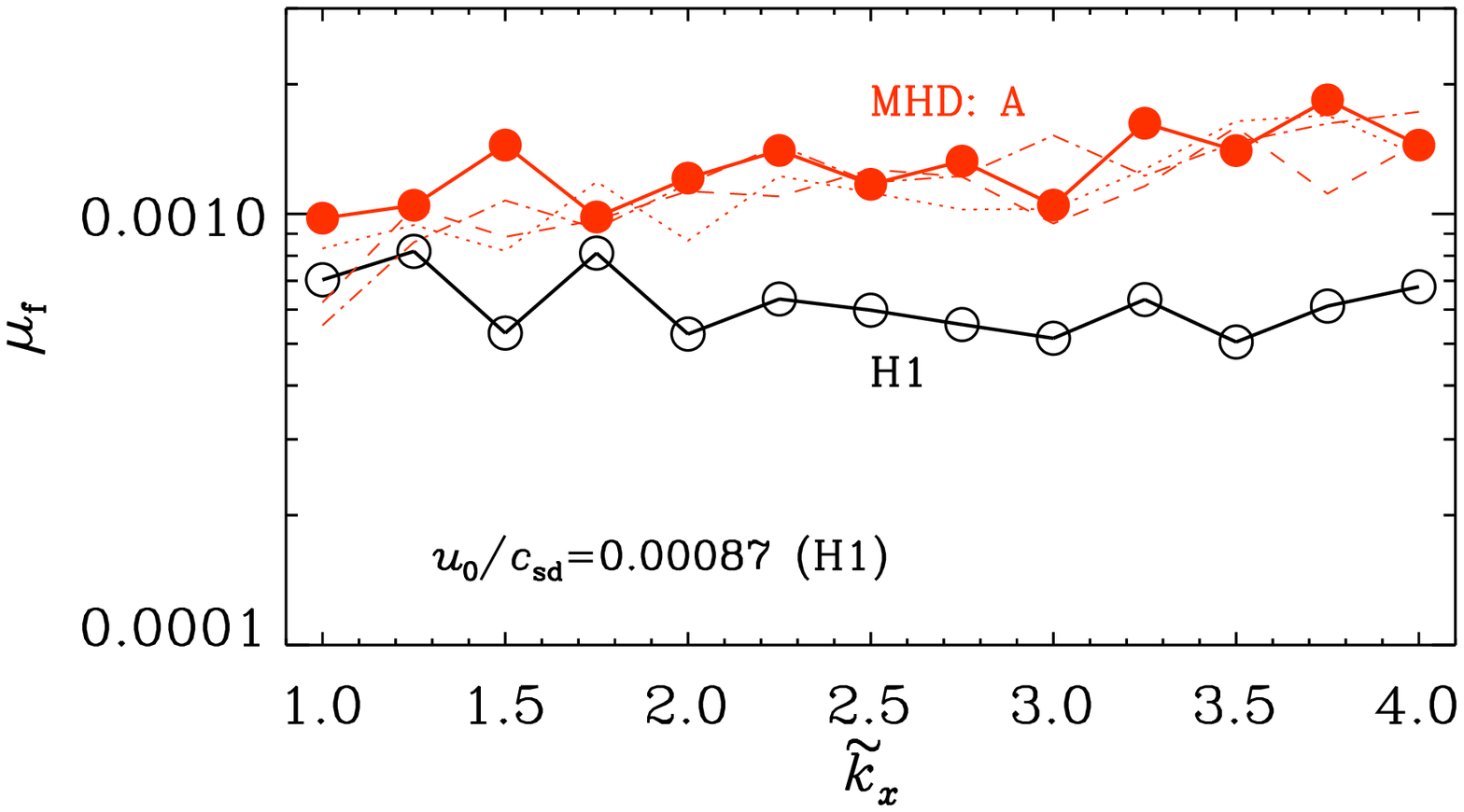}
\includegraphics[width=0.49\textwidth]{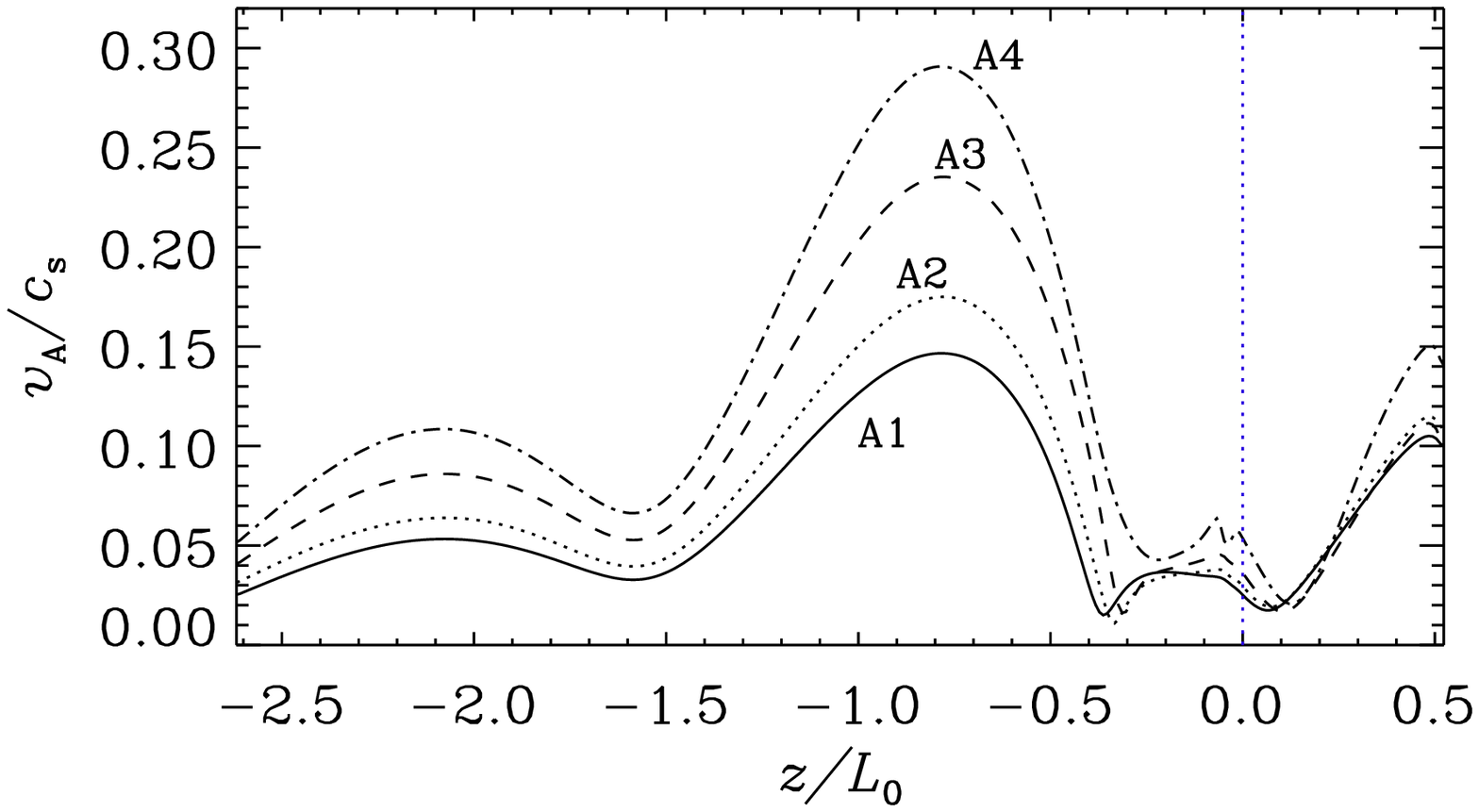}
\caption{Top: same as top panels in \Fig{hydro},
but for MHD runs A1 (left) and A4 (right).
Bottom left: 
normalised
mode strength $\mu_f$ of the $f$-mode as a
function of $\widetilde{k}_x$ for MHD runs A1--A4
(four overlapping red curves with filled circles for A1)
as well as for the hydrodynamic run H1
(open circles; solid black);
bottom right:
vertical profiles of $\vA/\cs$
in the saturated state of the background magnetic field
for models A1 (solid), A2 (dotted), A3 (dashed) and A4 (dash-dotted).
See \Tab{table_runs} for more details.
}\label{fmm_A}
\end{center}
\end{figure}

\begin{figure}
\begin{center}
\includegraphics[width=0.49\textwidth]{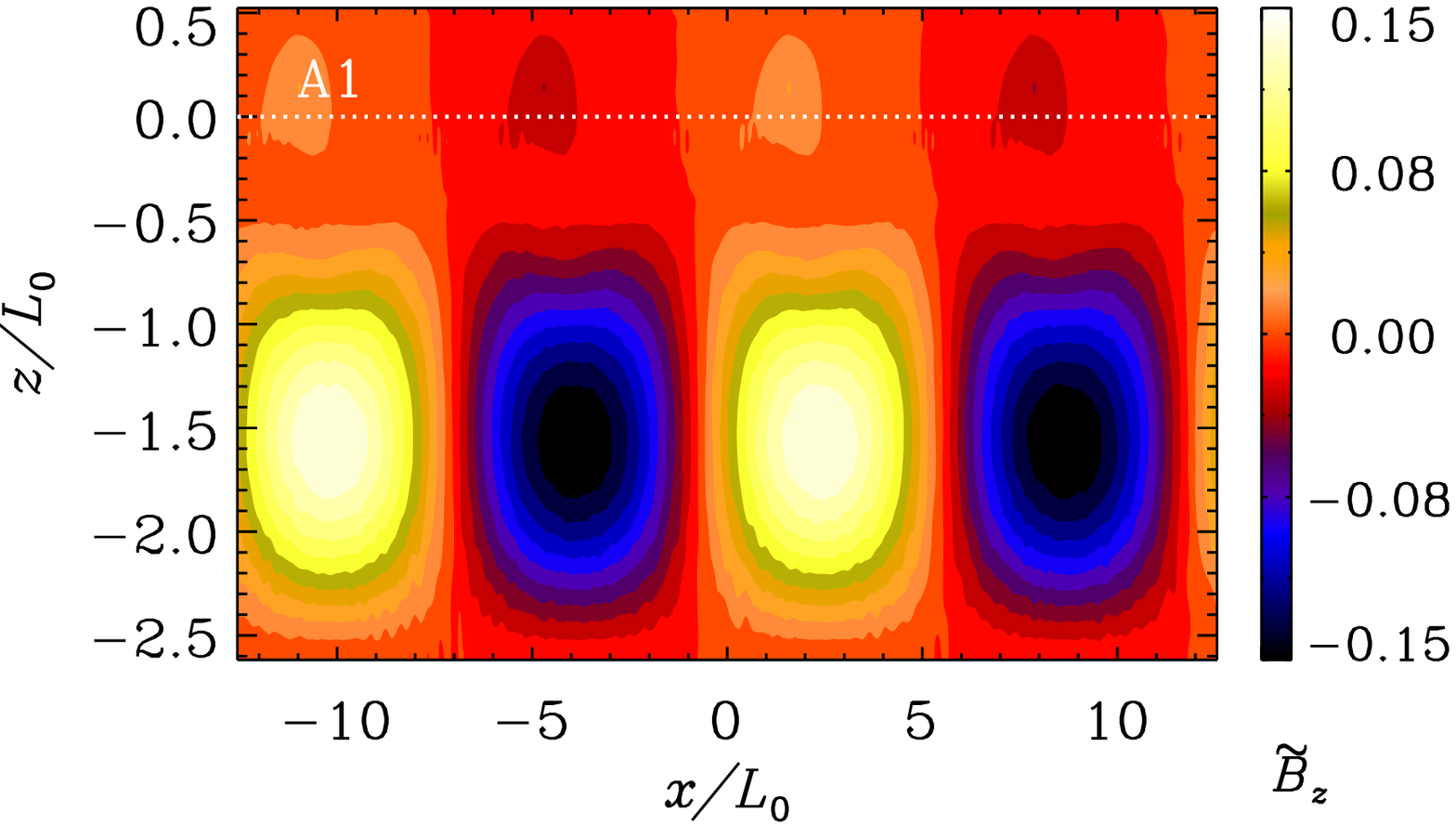}
\includegraphics[width=0.49\textwidth]{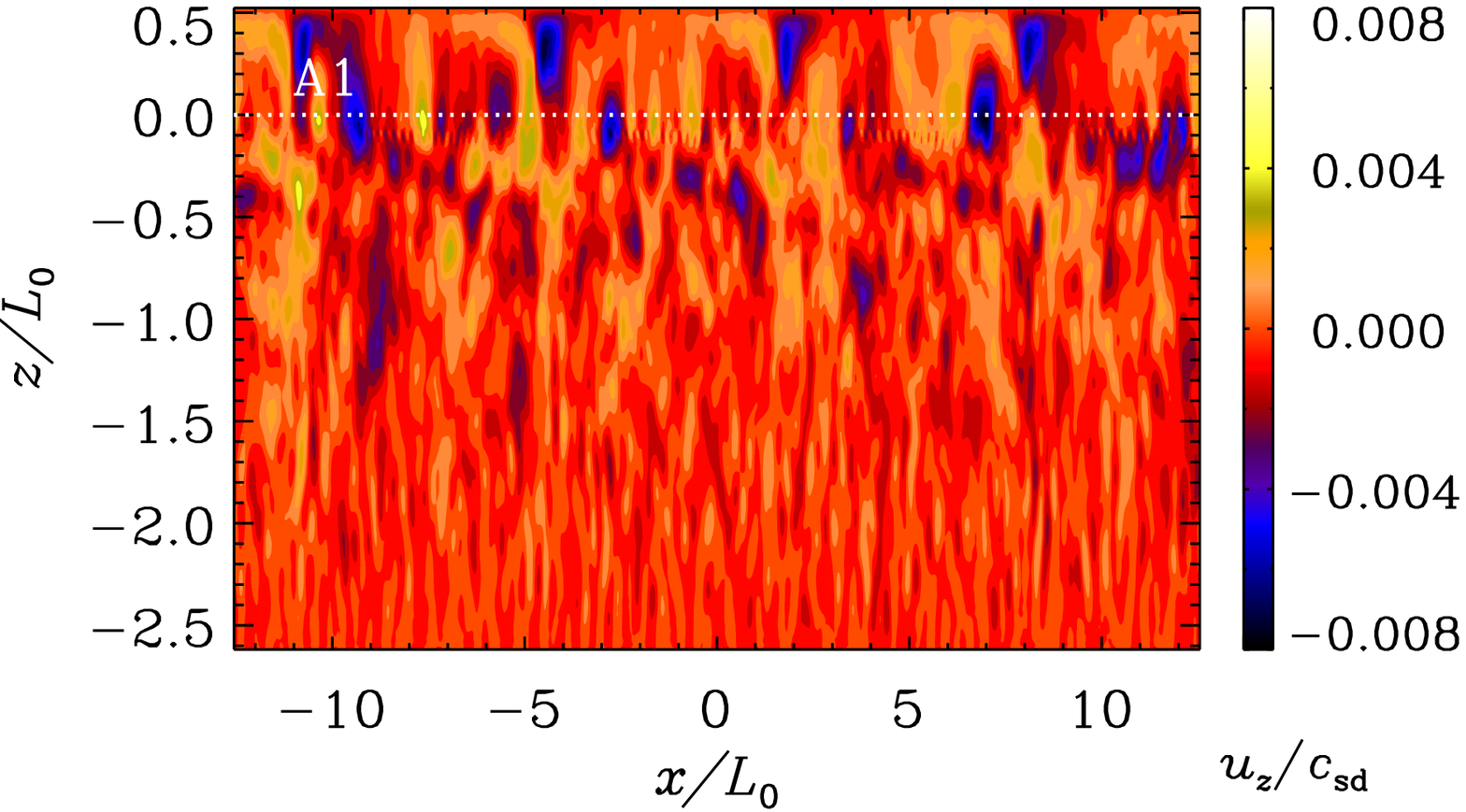}
\includegraphics[width=0.49\textwidth]{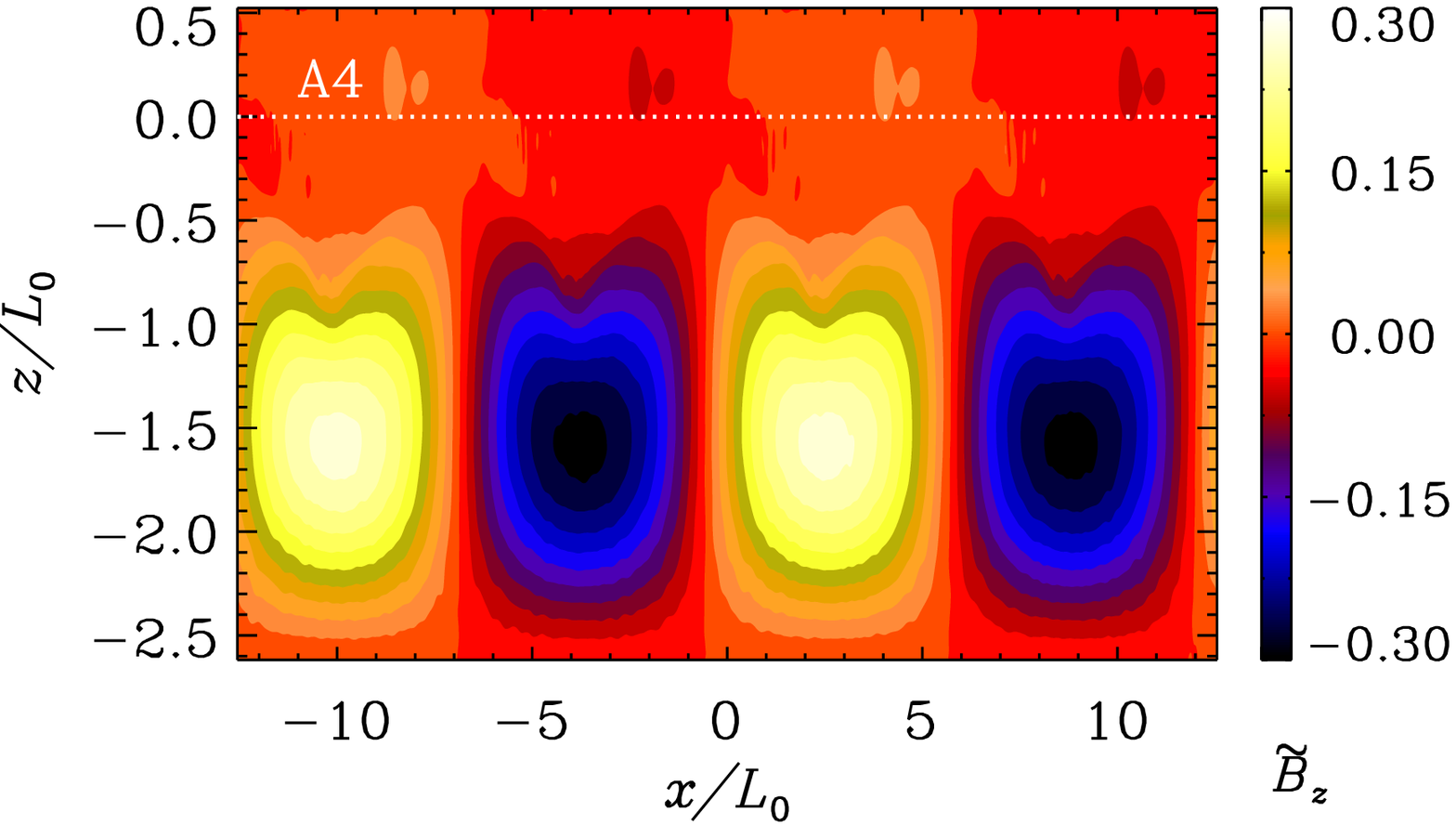}
\includegraphics[width=0.49\textwidth]{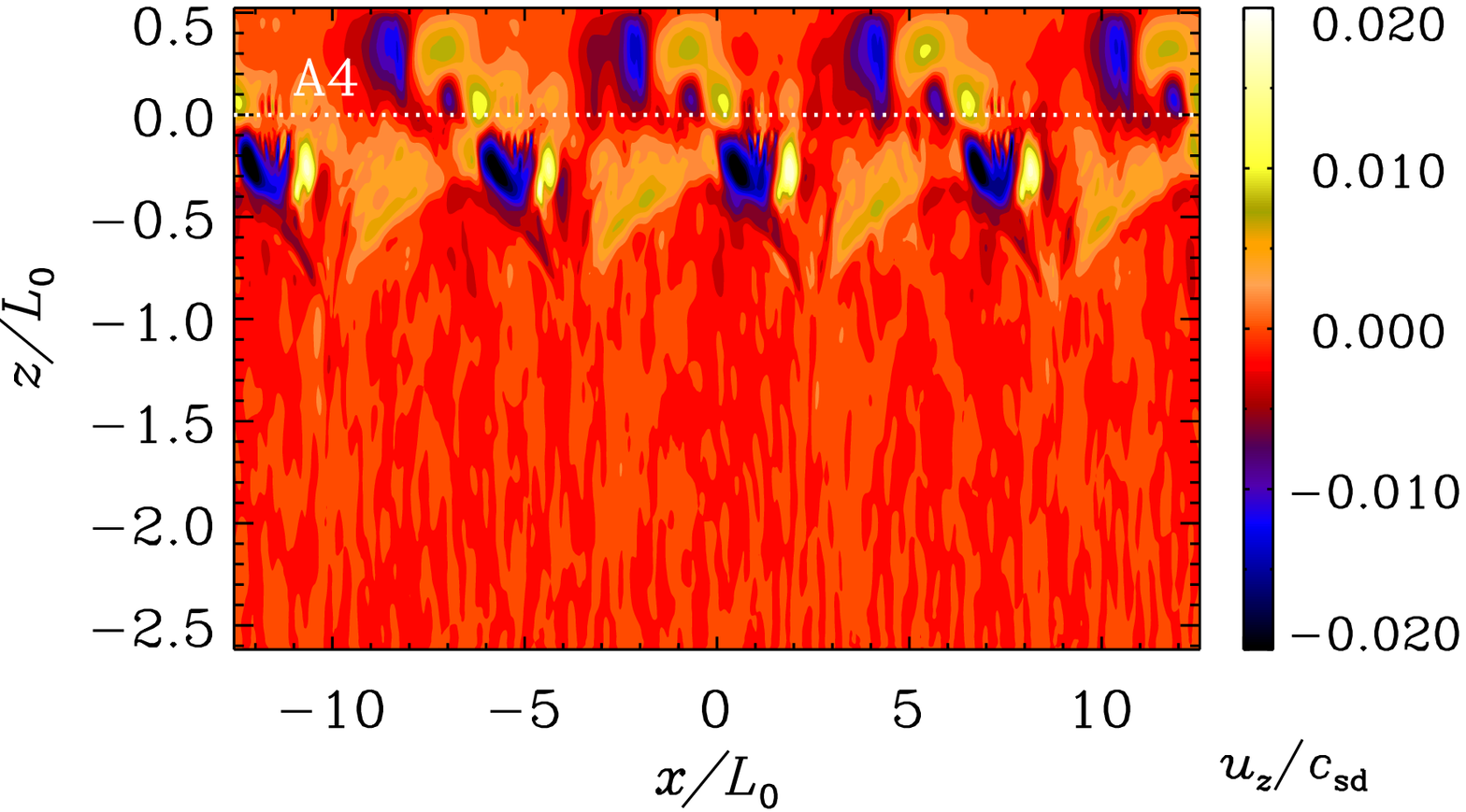}
\caption{Snapshots of normalised vertical magnetic
field $\tilde{B}_z=B_z/\csd\sqrt{\mu_0 \rd(0)}$
(left), and $u_z/\csd$ (right)
in the saturated state of the background magnetic field 
$\BB_0$ for MHD runs A1 (top) and A4 (bottom) shown in the $x$-$z$ plane
with dotted horizontal lines marking the location of the
interface at $z=0$.
}\label{snapshots_A}
\end{center}
\end{figure}

Here we investigate the effects of harmonically varying
subsurface magnetic fields, maintained by the EMF
$\emfA$ as given in \Eq{emfA}, on the $f$-mode.
We refer first to the $k\omega$ diagrams for runs A1 and A4,
shown in the top panels of \Fig{fmm_A}.
The vertical stripes at low frequencies correspond to the
unstable Bloch modes, which arise due to the periodicity of the
background magnetic field; see \citet{SBR14}.
Comparing these $k\omega$ diagrams with the one from
the suitable hydrodynamic run H1 (see \Tab{table_runs}) shown in
\Fig{hydro}, we see that the $f$-modes 
in the MHD runs are very different, thus already providing 
evidence that the $f$-mode is sensitive to subsurface magnetism
in our setup.
This, in a sense, confirms earlier analytical work, e.g.\ \cite{R81},
where larger frequencies of the $f$-mode are expected, although
the fanning and associated mode strengthening were not anticipated then.
Let us define the $z$-dependent rms Alfv\'en speed,
$v_{\rm A}(z)$, as
\EQ
v_{\rm A}(z) = \sqrt{\frac{\langle B^2 \rangle_x(z)}{\mu_0
\langle \rho \rangle_x(z)}}\,,
\label{vaz}
\EN
where $\bra{\cdot}_x$ denotes  averaging along the $x$ direction.
The vertical profiles of $v_{\rm A}(z)/\cs(z)$ are shown in
the bottom right panel of \Fig{fmm_A} for all A runs.

In the bottom left panel of \Fig{fmm_A}, we show the wavenumber dependence
of the 
normalised
mode strength $\mu_{\rm f}$ of the $f$-mode 
for the A runs and the corresponding hydrodynamic run H1.
All the input parameters for these runs are the same
(see \Tab{table_runs}), 
except for the magnetic field strength which is gradually
increased from A1 to A4 while keeping
its maximum at the same depth; see \Fig{fmm_A}.
Compared to the run H1, the mode strength is clearly
larger in the A runs for all $\widetilde{k}_x>1.5$. Interestingly, the 
mode strength
is nearly the same for runs A1--A4, while the line width
(or fanning) of the $f$-mode increases from A1 to A4, as
can be inferred from the $k\omega$ diagrams.
Snapshots of the vertical components
$B_z$ and $u_z$ are shown
in \Fig{snapshots_A} for models A1 and A4 in the
saturated stage of the background magnetic fields,
where their maxima lie more than one pressure scale height below the
interface at $z=0$.
While no large scale flow is noticeable in the weakly magnetised case A1,
we do see a systematic flow pattern in model A4 with a stronger
magnetic concentration. Despite the appearance of
systematic flows in these MHD runs, the normalised
$f$-mode strength is found to be nearly the same
in all the A runs, as can also be seen from the 
corresponding values of $\Sigma_{\rm f}$ 
(identical when rounded to three decimal places)
listed in \Tab{table_runs}.

\subsection{MHD B-type runs}

\begin{figure}
\begin{center}
\includegraphics[width=0.49\textwidth]{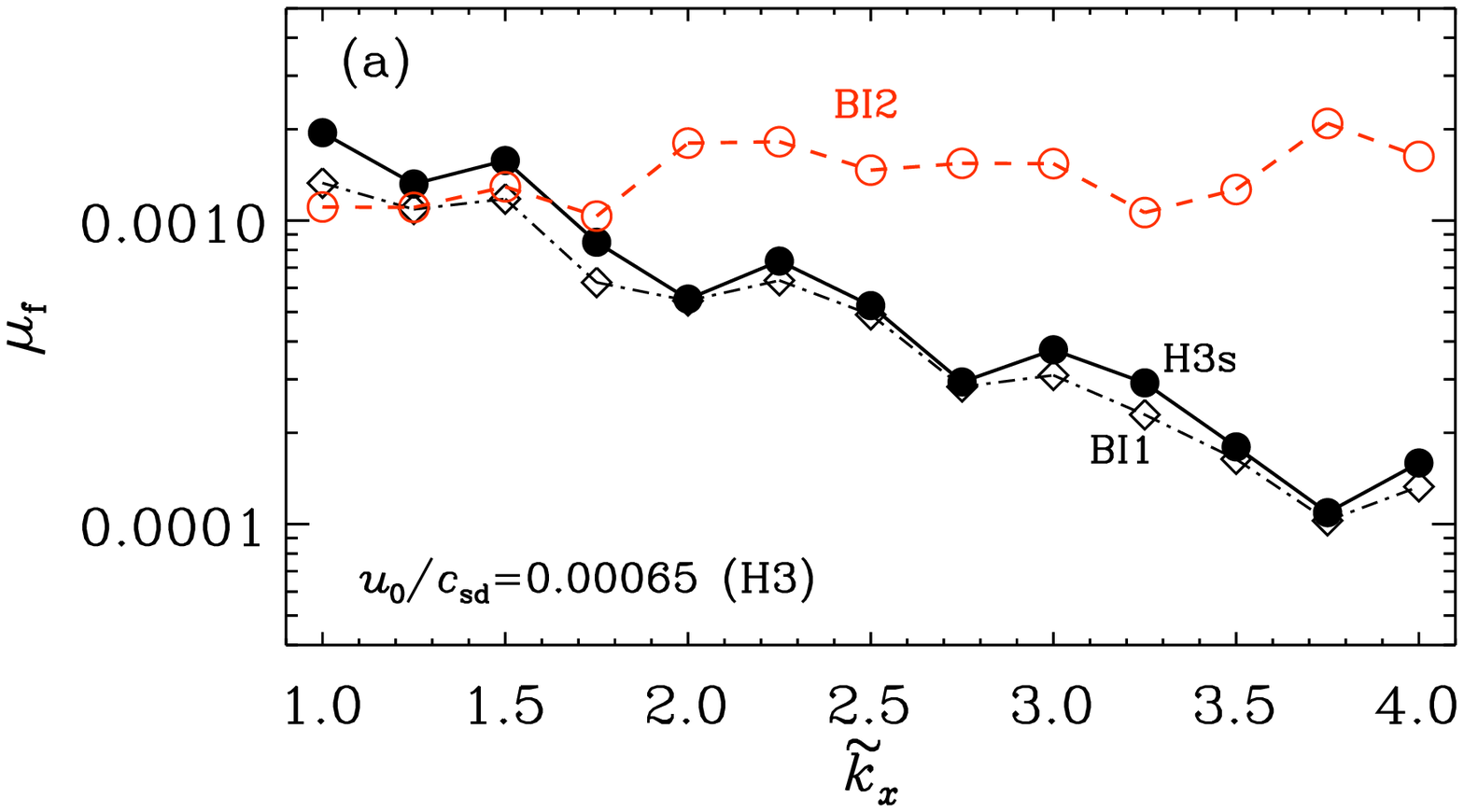}
\includegraphics[width=0.49\textwidth]{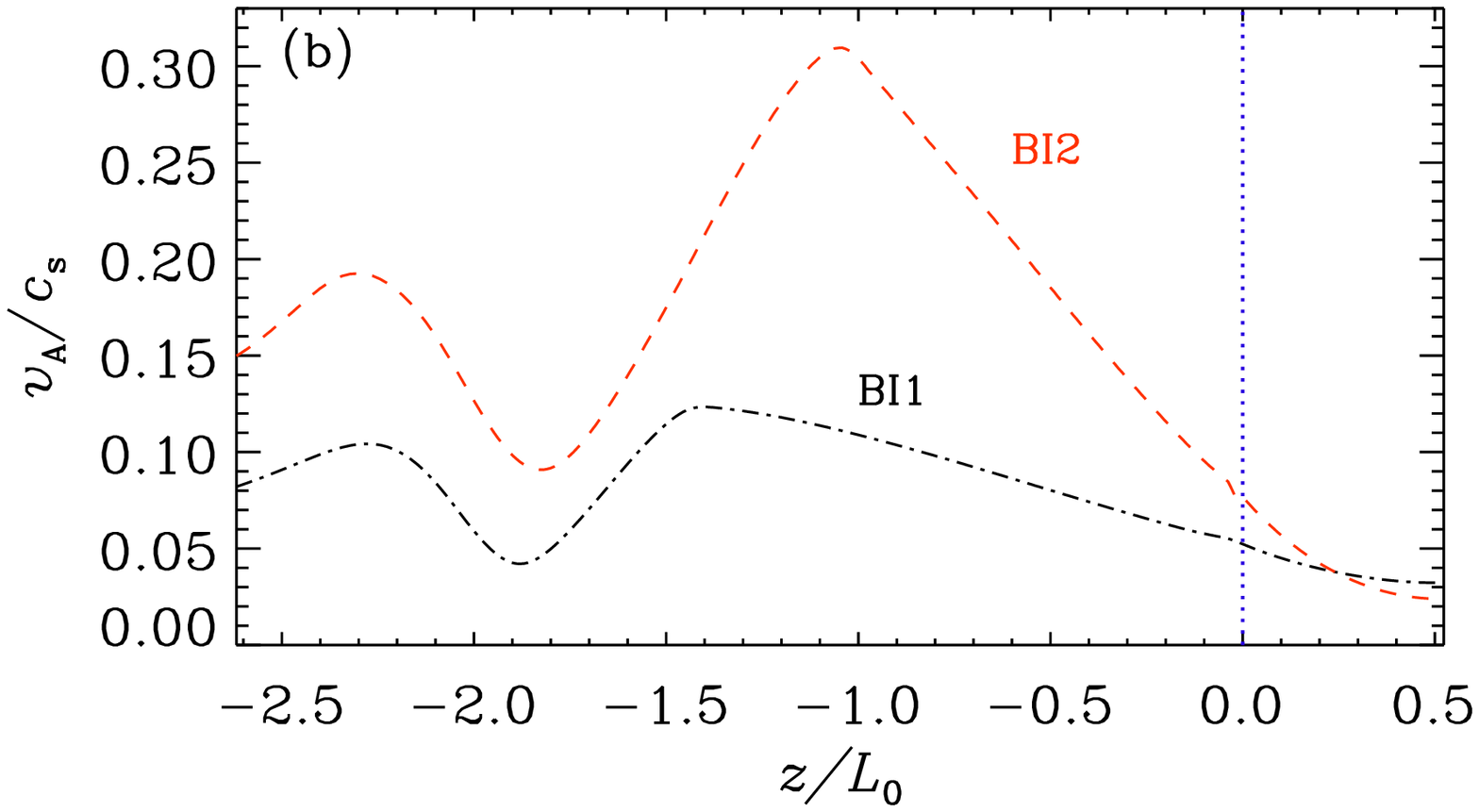}
\includegraphics[width=0.49\textwidth]{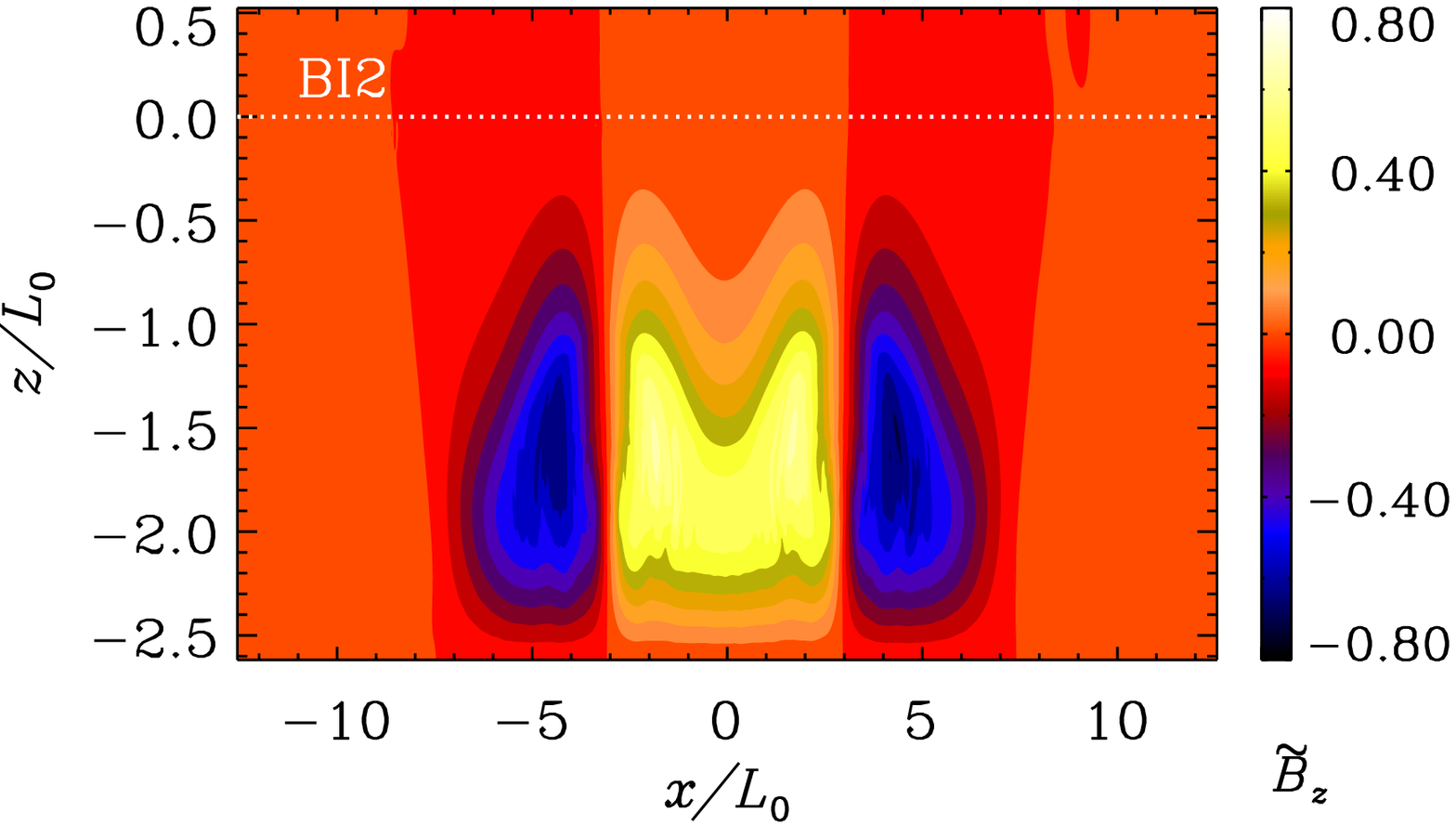}
\includegraphics[width=0.49\textwidth]{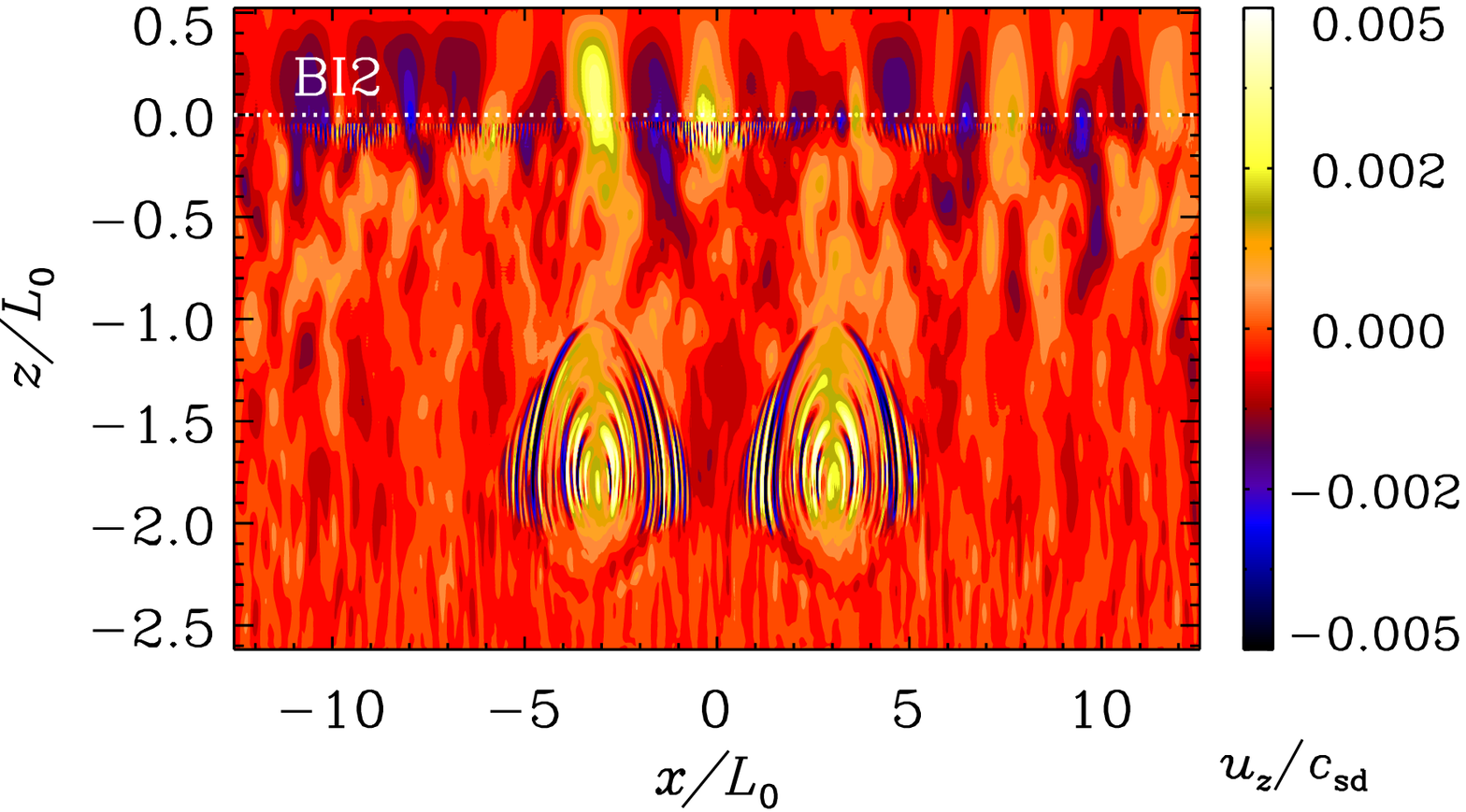}
\caption{Top: (a) normalised
mode strength $\mu_f$ of the $f$-mode as a
function of $\widetilde{k}_x$ for
MHD B-runs BI1 (diamonds; dash-dotted black) and BI2
(open circles; dashed red) as well as for the appropriate hydrodynamic run H3
(filled circles; solid black), and (b) vertical profiles of $\vA/\cs$
in the saturated state of the background magnetic field
for the corresponding MHD cases,
shown in left and right panels, respectively.
Bottom: same as top panels in \Fig{snapshots_A} but for the run BI2.
See \Tab{table_runs} for more details.
}\label{fmm_BI}
\end{center}
\end{figure}

Next we turn to MHD B-type models, where
{\it localised} magnetic concentrations
are maintained below the surface by the EMF $\emfB$, as given in \Eq{emfB}.
Their
horizontal extent 
$\Delta x_B=x_2-x_1$
is thus
restricted, with $\Delta\tilde{x}_B\approx4\pi$
(i.e., half the horizontal extent of the domain)
for the two BI-type models BI1 and BI2
and
$\Delta\tilde{x}_B\approx7\pi$ for model BII;
see the bottom left panels of \Figs{fmm_BI}{fmm_BII},
and \Tab{table_runs} for details.

\begin{figure}
\begin{center}
\includegraphics[width=0.49\textwidth]{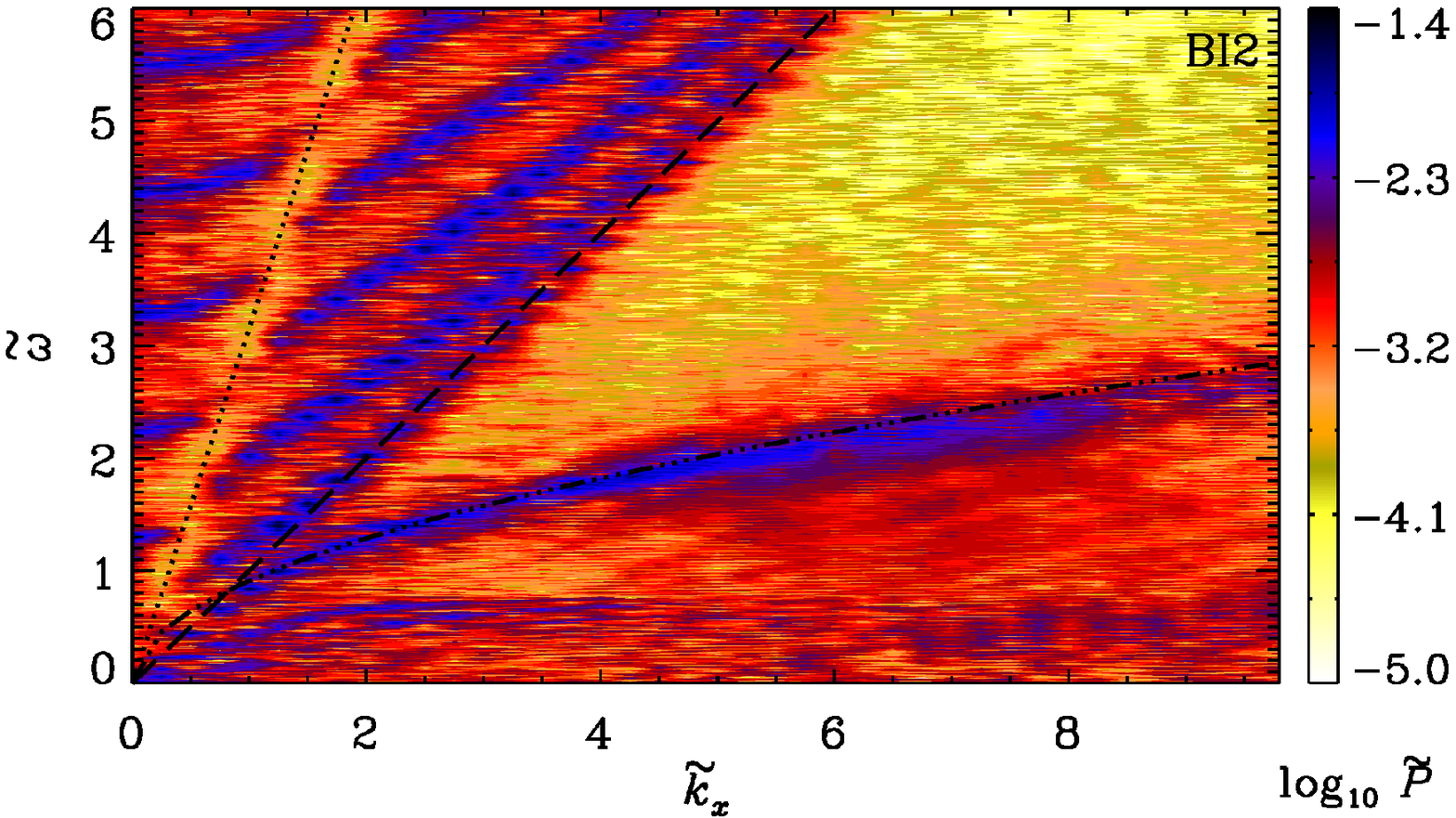}
\includegraphics[width=0.49\textwidth]{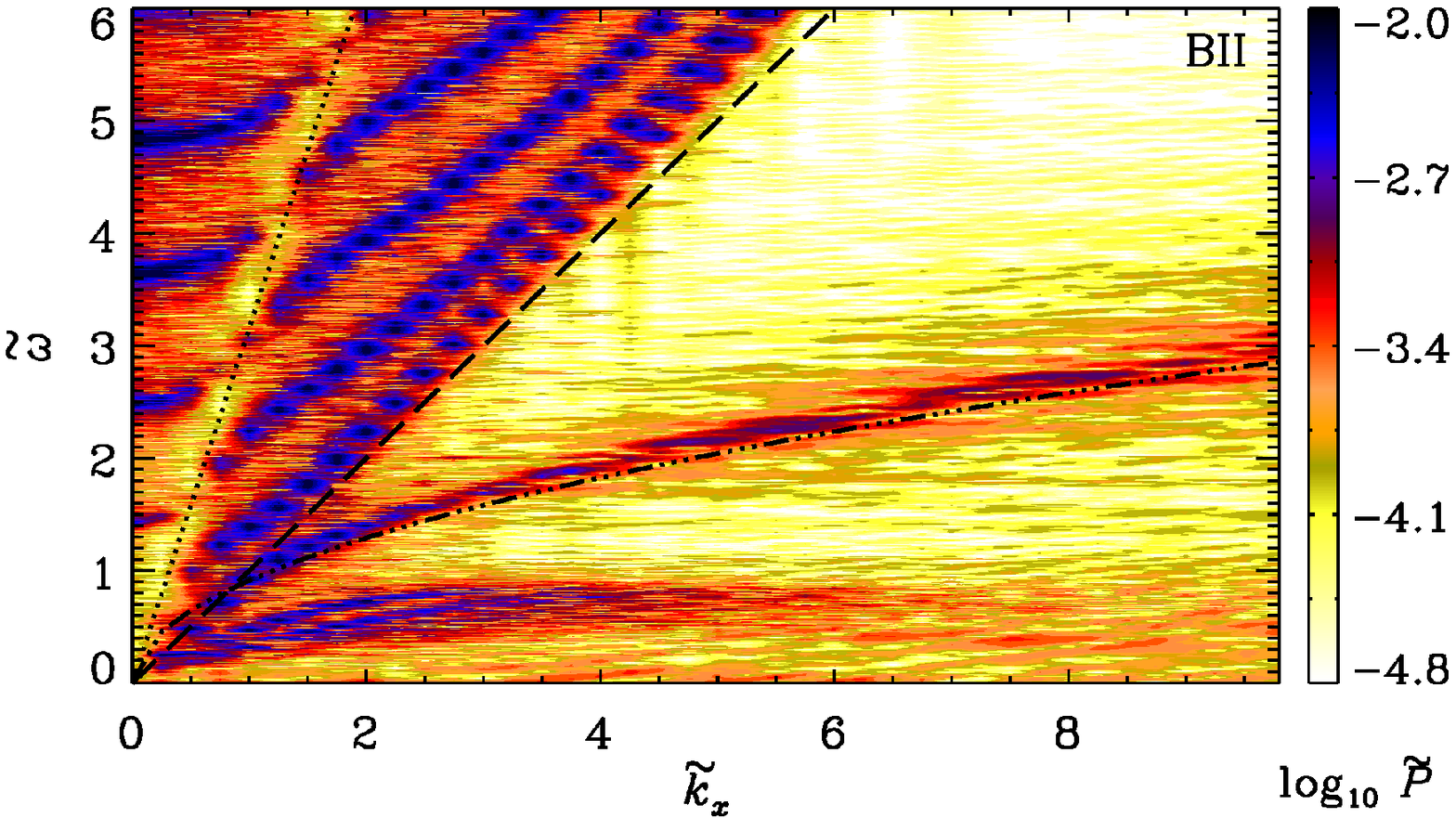}
\caption{$k\omega$ diagrams for MHD B-type runs BI2 (left) and
BII (right); see \Tab{table_runs} for more details.
}\label{ko_B}
\end{center}
\end{figure}

The $k\omega$ diagrams for models BI2 and BII are shown in
\Fig{ko_B}. Remarkably, the unstable Bloch modes
seen as vertical stripes at low frequencies 
in the A-type runs are completely absent here.
These were first found in \cite{SBR14} and thought
to arise due to the periodicity of the
background magnetic field \citep{BH87}.
Given that the appearance of solar ARs during most phases of the magnetic
activity cycle is hardly ever periodic in
longitude, the Bloch modes are unlikely to be seen. Yet,
during very active phases of the Sun, these modes are potentially observable 
in the $k\omega$ or ring diagrams that
are constructed from sufficiently large patches covering many
ARs 
which show
a regular pattern of alternating positive and negative
vertical magnetic field along the azimuthal direction.
On the other hand, given the drastic difference between our A- and B-type
runs, we must acknowledge that the Bloch modes are rather sensitive to the
exact matching of the wavelength and the size of the patch.
This limits the observability of Bloch modes severely.

As for models A1--A4, a direct comparison of diagnostic diagrams
shown in \Figs{hydro}{ko_B} reveals that the $f$-mode is
significantly perturbed in the presence of subsurface magnetic
fields; compare, e.g., BI2 with the corresponding hydrodynamic case H3
where the hydrodynamic forcing is restricted to
$\tilde{z}<0$ (see also \Tab{table_runs}).
The $f$-mode in the MHD run BI2 appears to exist
at all wavenumbers shown, whereas it is markedly damped
at large wavenumbers in the corresponding hydrodynamic run H3.
Moreover, the fanning of the $f$-mode, as discussed above
\citep[see also][]{SBR14}, is much weaker
here for localised magnetic field concentrations.

The wavenumber dependence of $\mu_{\rm f}$ for the two BI-runs
and the relevant hydrodynamic run H3, as well as the vertical profiles
of the normalised Alfv\'en speed for these MHD runs are shown in the top
panels of \Fig{fmm_BI}. For a meaningful comparison, the $f$-mode
strengths were determined from data sets
with identical time-spans, which can, in general, affect
the mode strengths. The label H3s, 
used
in \Figp{fmm_BI}{a} signifies that the data are from run H3, but
$\mu_{\rm f}$ is determined from a
shorter time-series, matching the ones in BI1 and BI2.

\begin{figure}
\begin{center}
\includegraphics[width=0.49\textwidth]{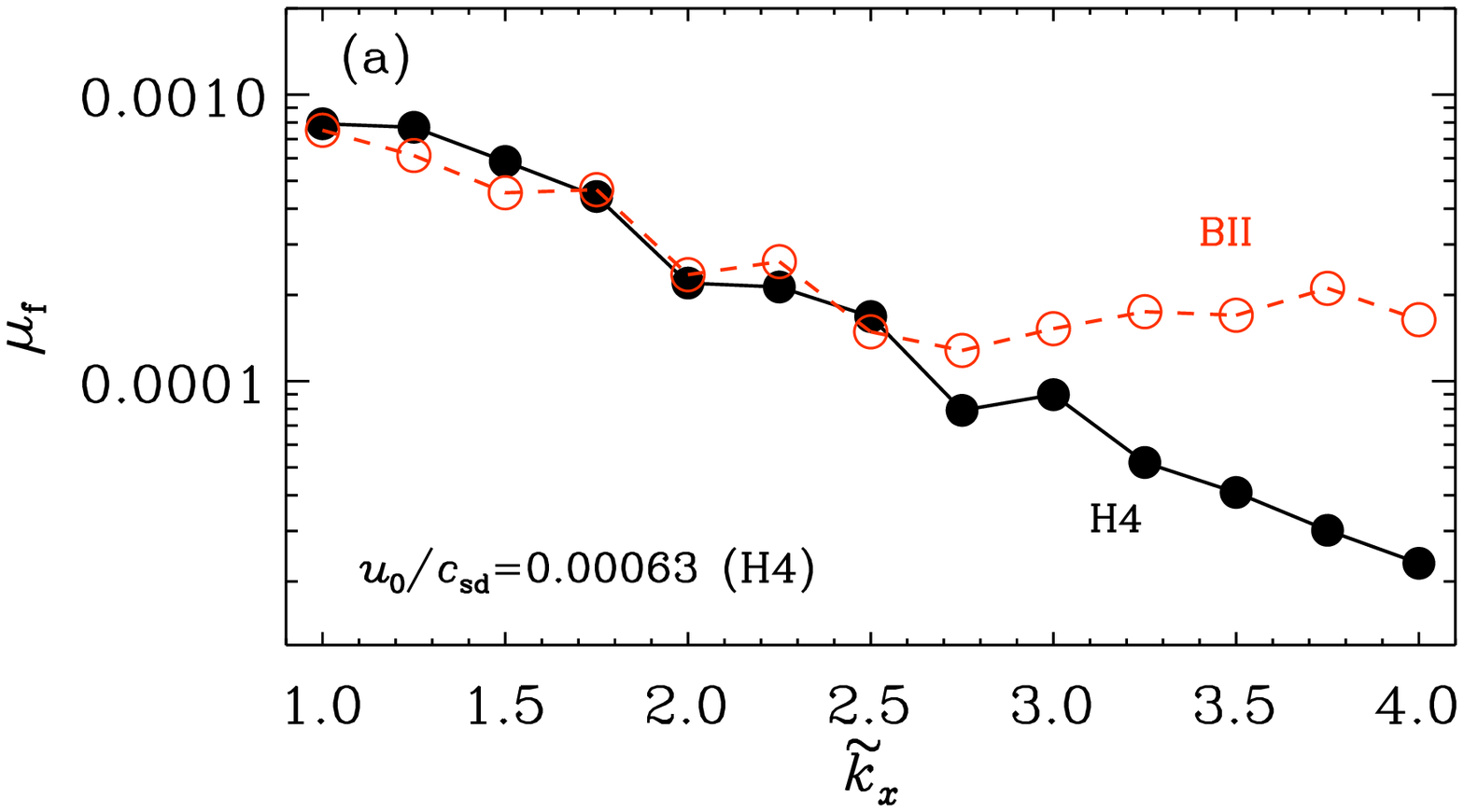}
\includegraphics[width=0.49\textwidth]{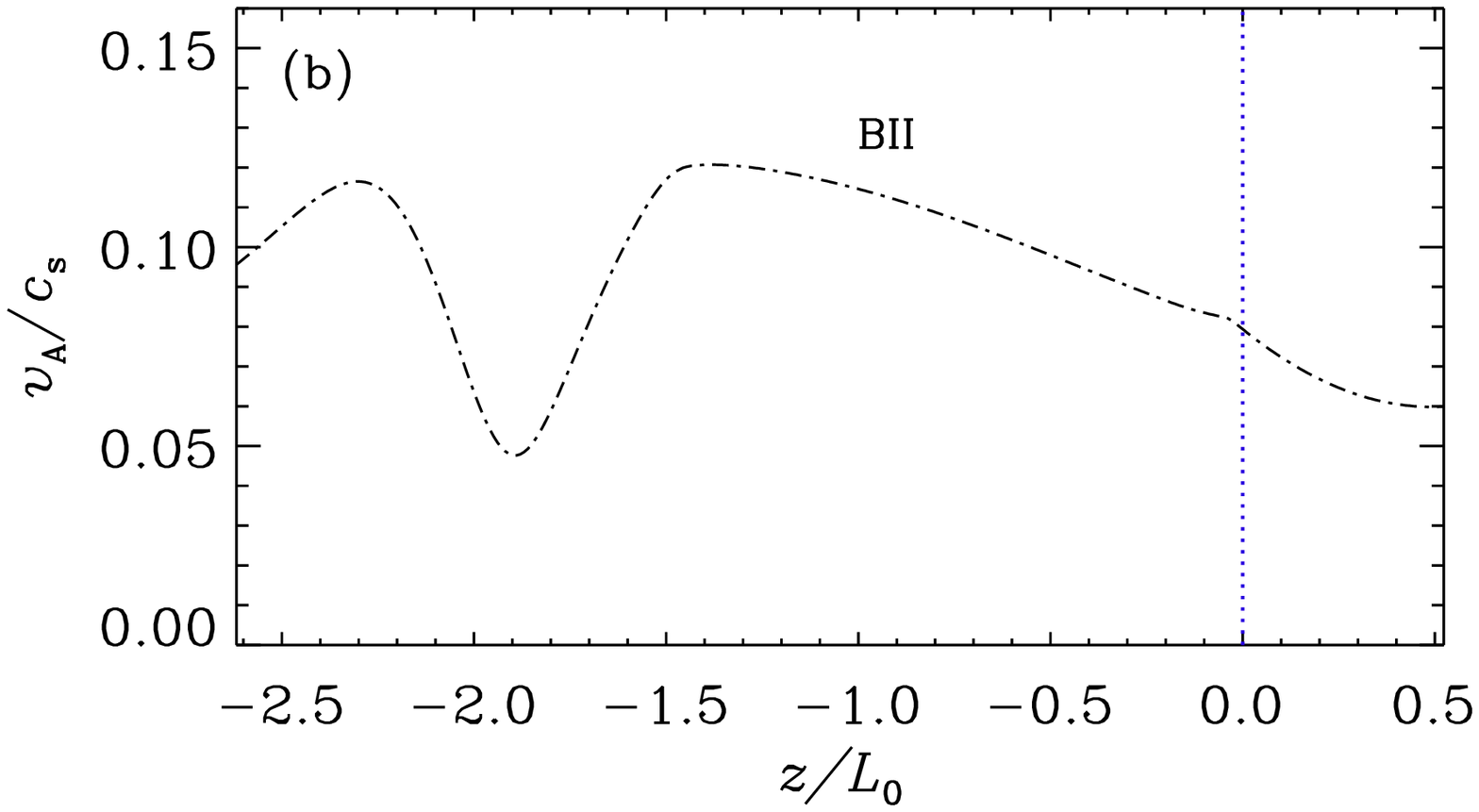}
\includegraphics[width=0.49\textwidth]{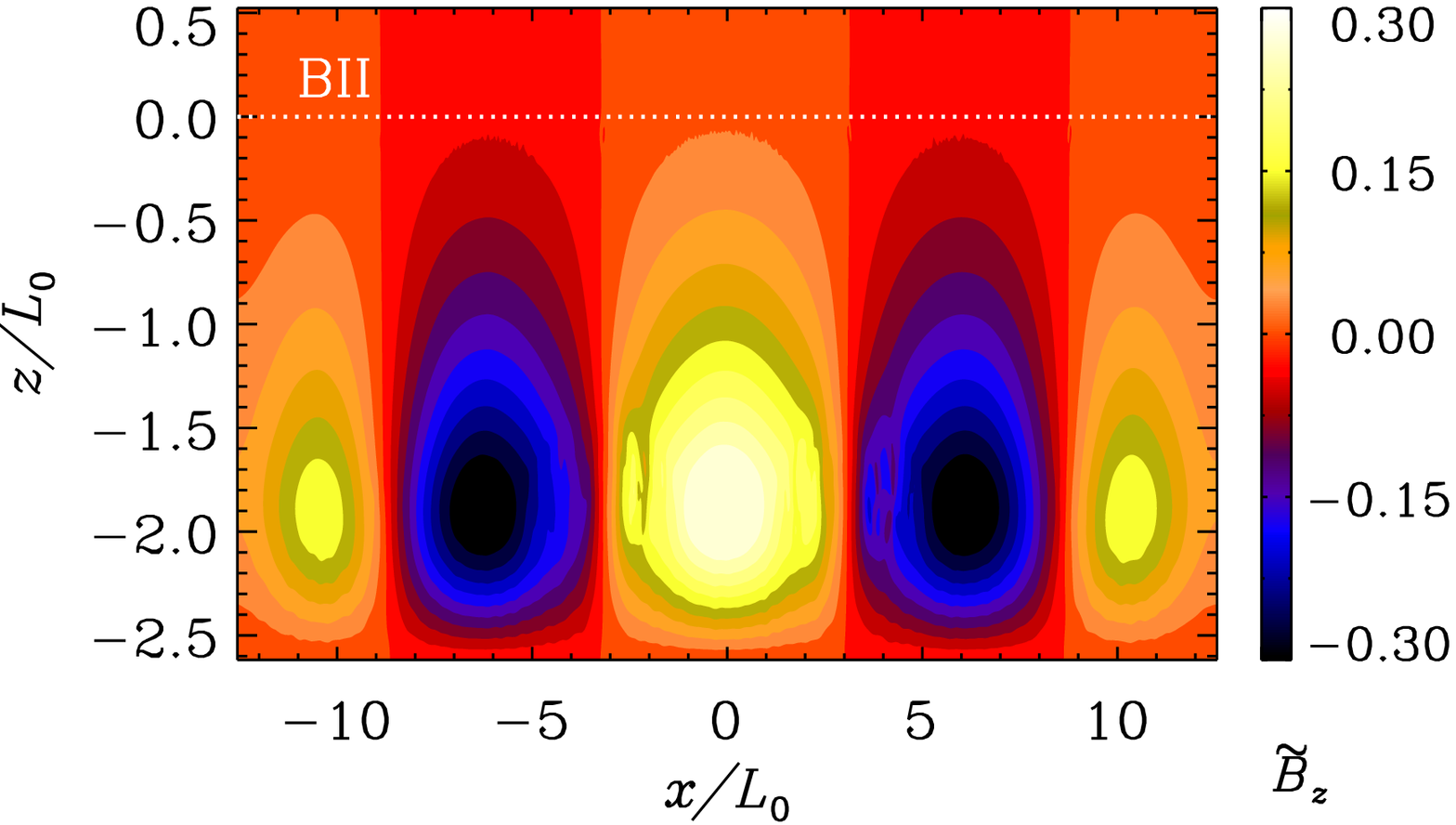}
\includegraphics[width=0.49\textwidth]{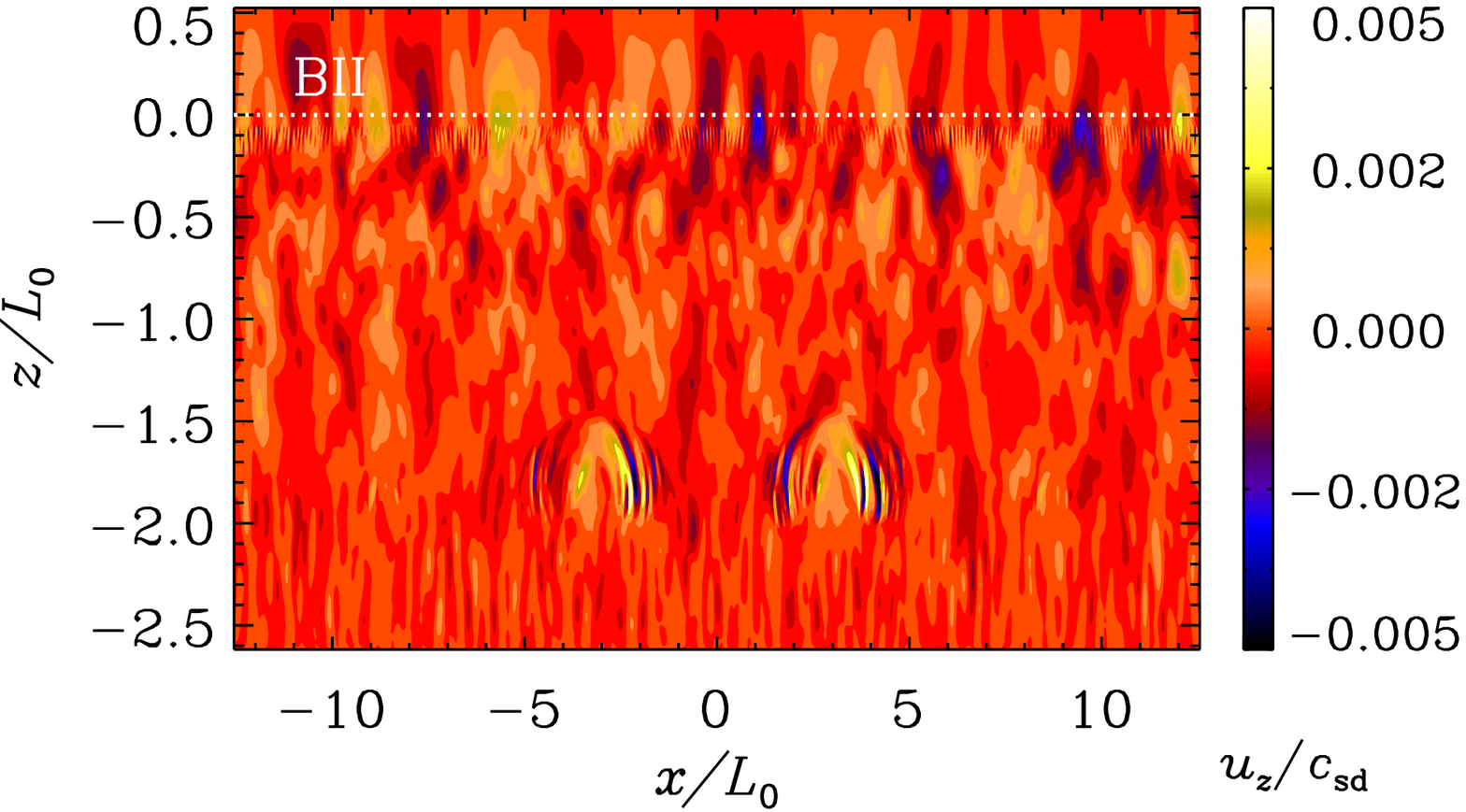}
\caption{Same as \Fig{fmm_BI} but for the MHD run BII;
see \Tab{table_runs} for more details.
}\label{fmm_BII}
\end{center}
\end{figure}

From \Figp{fmm_BI}{a} we see that $\mu_{\rm f}$ for the MHD run
BI1 with weak magnetic concentration is nearly the same
as in case of run H3s. Note also that the maximum
of the Alfv\'en-to-sound speed ratio in this case lies at a
larger depth compared to the A runs; see \Figp{fmm_BI}{b}
and \Tab{table_runs}.
This may be another reason why we do not find mode
strengthening in BI1. However, when the magnetic
field is increased and its maximum moves upwards to somewhat
shallower depths as in case of BI2, we immediately find the
$f$-mode strengthening at wavenumbers $\widetilde{k}_x\geq1.8$.
Snapshots of $B_z$ and $u_z$ are shown in the bottom panels of
\Fig{fmm_BI} for model BI2.
The horizontal oscillations in $u_z$ in the proximity of regions of strong
magnetic field may seem to be numerical artifacts, but they do not look like
the regular ``ringing'' one encounters when a simulation is underresolved.
As they are resolved by a few mesh points
per period, they may well be real.
Comparing these with the relevant hydrodynamic run H3
(see \Fig{hydro}) we find that the vertical motions are
comparable and have no large scale feature at the interface.
As expected, a systematic flow pattern is indeed seen in
deeper layers around the maximum of the magnetic concentration,
but, as was argued before based on the A runs, it is not clear
whether such flows can affect the mode strength of the $f$-mode
which is determined from vertical motions only at the interface.

In those of our experiments where the forcing is further
restricted to layers
$\tilde{z}<-0.2$, as in the MHD model BII or in the hydrodynamic run H4
(see \Tab{table_runs}), we still find evidence of $f$-mode
strengthening in the presence of subsurface magnetic fields, but
at larger wavenumbers (in this case for $\widetilde{k}_x\geq2.5$).
This is demonstrated in \Fig{fmm_BII}, where snapshots of
$B_z$ and $u_z$ are also shown.
In both models, $\mu_{\rm f}$ first decreases with
$\widetilde{k}_x$, and, while it saturates at a constant level
beyond about $2.5$ in case of BII, it shows a monotonic decrement
in H4; see \Figp{fmm_BII}{a}.
Interestingly, the $f$-mode strengthening is 
thus discernible in model BII, but is not present
in model BI1, although both have similar magnetic field concentrations
at nearly the same depth;
see \Tab{table_runs}, and compare panel~(b) of \Figs{fmm_BI}{fmm_BII}.
This is likely due to the larger horizontal extent
of the magnetic structure in the case of BII as compared to BI1.

Understanding the exact physical mechanism, responsible for the
strengthening of the $f$-mode observed here, is beyond the
scope of the present paper.
Both the imposed magnetic field
and the systematic flow driven by it, can, in principle,
influence frequency and
strength of the $f$ mode.
For assessing the relative importance of these two factors
in our studies,
let us compare Runs~H1 and A1.
As can be seen from the upper right panel of \Fig{snapshots_A},
the systematic flow can hardly be seen near the interface,
mostly because of the magnetic field being weak in
case A1.
So we might conclude that the considerable
increase in the relative mode amplitude $\Sigma_f$
is mainly due to the magnetic field.
By contrast, Run~A4 (\Fig{snapshots_A}, lower right panel)
exhibits a systematic flow pattern dominating over the random component
of the flow.
Hence, it cannot be decided to what extent the systematic flow is
affecting $\Sigma_f$.
In Runs~A1-A4, $\Sigma_f$ is nearly the same, although
the magnetic field is changing by a factor of two. As the magnetic
field alone seems to strengthen the $f$ mode, 
this could be an indication of a counteracting effect of the
systematic flow, which grows with the magnetic field, too.
More work is needed to understand the role of a systematic or
mean flow component on the strength of the $f$-mode.
This will be the focus of a separate investigation.

\section{Conclusions}
\label{conc}

We have shown numerically that the $f$-mode is significantly perturbed
in the presence of subsurface magnetic fields concentrated one to a few scale
heights below the photosphere. The fanning (or increase in line width)
of the $f$-mode was first reported in \cite{SBR14} based on
harmonically varying magnetic fields, but the associated
mode strengthening was not emphasised there.
Here we extended this by also investigating the effects of
localised bipolar magnetic fields, resembling more realistically
the active region precursors.
The fanning effect of the $f$-mode is found to be weaker in the case of
localised magnetic field concentrations.
Motivated by the observational findings of \cite{SRB16},
where the strengthening at large wavenumbers of the $f$-mode
prior to the emergence of about
half a dozen active regions, we focus here primarily on the
phenomenon of $f$-mode strengthening.
In our numerical investigations reported here, the $f$-mode strength
is found to be clearly larger at high horizontal
wavenumbers in a variety of magnetic background states,
compared to their corresponding hydrodynamic cases.
While the fanning
of the $f$-mode at large $k_x$ is more sensitive to the
strength of the magnetic field at a given depth, the mode strength
itself shows a dependence on the location of the magnetic
concentration beneath the
surface. Thus, it is indeed remarkable that the properties of the $f$-mode
trace so effectively both the location and the strength of 
subsurface magnetic fields that are not yet directly visible at the
photosphere.

We note that damping of the $f$-mode due to its resonant
coupling with atmospheric magnetic fields or due to absorption caused
by sunspots is known \citep{CB97,E06,PEG07}, and also seen
in observational studies of \cite{SRB16} \emph{after} the emergence of
active regions. But, the strengthening of the $f$-mode
due to magnetic fields confined below the photosphere is a qualitatively
different finding, which is numerically confirmed in the present work.

Let us now estimate the instrumental requirements to robustly
probe and detect solar $f$-mode perturbations of the
kind explored in this paper.
In order to detect both fanning and mode-strength gain,
according to \Figs{fmm_A}{fmm_BI},
$\widetilde{k}_x\gtrsim2$ needs to be reached.
With $\gamma \Hd \approx 0.5\Mm$ and $R_\odot=700\Mm$ being the
solar radius, 
this corresponds to spherical harmonic degrees $\ell \gtrsim 2800$.
However, indications of $f$-mode strengthening prior to 
active region formation could indeed be available even at
somewhat smaller degrees, as was seen in the observational work of
\cite{SRB16} based on data from the HMI instrument, whose
sensitivity drops beyond $\ell\approx2000$. The precursor
signal is expected to improve at larger wavenumbers and therefore
higher resolution data from other facilities and upcoming missions
will be critical not only for establishing the connection between
the solar $f$-mode and the subsurface magnetic fields, but 
also for providing
deeper insights into the sunspot formation mechanism, which is
still an outstanding topic in solar physics.
More work is needed to understand the physical mechanism responsible
for such a strengthening of the $f$-mode and this will be
attempted elsewhere.
In particular, it would be useful to extend these models
to explore effects of 3-dimensional magnetic configurations;
see the review of \cite{E06} for possible effects on
mode damping and mode conversion.

\section*{Acknowledgements}

We thank the two referees for useful comments on the paper.
We acknowledge the allocation of computing resources provided by the
CSC -- IT Center for Science Ltd.\ in Espoo, Finland, which is
administered by the Finnish Ministry of Education (project 2000403),
and by the Swedish National Allocations Committee at the Center for
Parallel Computers at the Royal Institute of Technology in Stockholm.
Financial support
from the Max Planck Princeton Centre for Plasma Physics (NS), Deutsche
Forschungsgemeinschaft Heisenberg programme (grant No.\ KA 4825/1-1;
PJK), the Academy of Finland ReSoLVE Center of Excellence (grant
No.\ 307411; MJK, MR, PJK), the National Science Foundation
Astrophysics and Astronomy Grant Program grant AST1615100,
and the University of Colorado through its support of the
George Ellery Hale visiting faculty appointment, is acknowledged. 


\end{document}